\documentclass[iop,apj,tighten,twocolappendix,numberedappendix]{emulateapj}
\usepackage{natbib}
\usepackage{amsmath,amssymb}
\usepackage{apjfonts}
\usepackage{xspace}
\usepackage[maxfloats=50]{morefloats}
\usepackage[plainpages=false, colorlinks=true, anchorcolor=blue, linkcolor=blue, citecolor=blue, bookmarks=false]{hyperref}
\usepackage{graphicx}
\shorttitle{NANOGrav Nine-year Astrometry}
\shortauthors{Matthews et al.}

\def\be{\begin{equation}}
\def\ee{\end{equation}}

\newcommand{\bb}{\begin{bmatrix}}
\newcommand{\eb}{\end{bmatrix}}

\newcommand{\kms}{km~s\ensuremath{^{-1}}}

\begin{document}

\title{The NANOGrav Nine-year Data Set: Astrometric Measurements of 37 Millisecond Pulsars}

\author{Allison~M.~Matthews\altaffilmark{1,2},
David~J.~Nice\altaffilmark{1},
Emmanuel~Fonseca\altaffilmark{3},
Zaven~Arzoumanian\altaffilmark{4},
Kathryn~Crowter\altaffilmark{3},
Paul~B.~Demorest\altaffilmark{5},
Timothy~Dolch\altaffilmark{6,7},
Justin~A.~Ellis\altaffilmark{8,17},
Robert~D.~Ferdman\altaffilmark{9},
Marjorie~E.~Gonzalez\altaffilmark{3,10},
Glenn~Jones\altaffilmark{11},
Megan~L.~Jones\altaffilmark{12},
Michael~T.~Lam\altaffilmark{6},
Lina~Levin\altaffilmark{12,13},
Maura~A.~McLaughlin\altaffilmark{12},
Timothy~T.~Pennucci\altaffilmark{2},
Scott~M.~Ransom\altaffilmark{14},
Ingrid~H.~Stairs\altaffilmark{3,9},
Kevin~Stovall\altaffilmark{15},
Joseph~K.~Swiggum\altaffilmark{12},
and
Weiwei~Zhu\altaffilmark{8,16}\\
}

\altaffiltext{1}{Department of Physics, Lafayette College, Easton, PA 18042, USA}
\altaffiltext{2}{University of Virginia, Department of Astronomy, P.~O.~Box 400325 Charlottesville, VA 22904-4325, USA}
\altaffiltext{3}{Department of Physics and Astronomy, University of British Columbia, 6224 Agricultural Road, Vancouver, BC V6T 1Z1, Canada}
\altaffiltext{4}{Center for Research and Exploration in Space Science and Technology and X-Ray Astrophysics Laboratory, NASA Goddard Space Flight Center, Code 662, Greenbelt, MD 20771, USA}
\altaffiltext{5}{National Radio Astronomy Observatory, P.~O.~Box 0, Socorro, NM, 87801, USA}
\altaffiltext{6}{Department of Astronomy, Cornell University, Ithaca, NY 14853, USA}
\altaffiltext{7}{Department of Physics, Hillsdale College, 33 E. College Street, Hillsdale, Michigan 49242, USA}
\altaffiltext{8}{Jet Propulsion Laboratory, California Institute of Technology, 4800 Oak Grove Dr. Pasadena CA, 91109, USA}
\altaffiltext{9}{Department of Physics, McGill University, 3600 rue Universite, Montreal, QC H3A 2T8, Canada}
\altaffiltext{10}{Department of Nuclear Medicine, Vancouver Coastal Health Authority, Vancouver, BC V5Z 1M9, Canada}
\altaffiltext{11}{Department of Physics, Columbia University, 550 W. 120th St. New York, NY 10027, USA}
\altaffiltext{12}{Department of Physics, West Virginia University, P.O.  Box 6315, Morgantown, WV 26505, USA}
\altaffiltext{13}{Jodrell Bank Centre for Astrophysics, School of Physics and Astronomy, The University of Manchester, Manchester M13 9PL, UK}
\altaffiltext{14}{National Radio Astronomy Observatory, 520 Edgemont Road, Charlottesville, VA 22903, USA}
\altaffiltext{15}{Department of Physics and Astronomy, University of New Mexico, Albuquerque, NM, 87131, USA}
\altaffiltext{16}{Max-Planck-Institut f{\" u}r Radioastronomie, Auf dem H{\" u}gel 69, D- 53121, Bonn, Germany}
\altaffiltext{17}{Einstein Fellow}

\begin{abstract}

Using the nine-year radio-pulsar timing data set from the North American Nanohertz Observatory for Gravitational Waves (NANOGrav), collected at Arecibo Observatory and the Green Bank Telescope, we have measured the positions, proper motions, and parallaxes for 37 millisecond pulsars. We report twelve significant parallax measurements and distance measurements, and eighteen lower limits on distance. We compare these measurements to distances predicted by the NE2001 interstellar electron density model and find them to be in general agreement. We use measured orbital-decay rates and spin-down rates to confirm two of the parallax distances and to place distance upper limits on other sources; these distance limits agree with the parallax distances with one exception, PSR~J1024$-$0719, which we discuss at length.  Using the proper motions of the 37 NANOGrav pulsars in combination with other published measurements, we calculate the velocity dispersion of the millisecond pulsar population in Galactocentric coordinates.  We find the radial, azimuthal, and perpendicular dispersions to be $46$, $40$, and $24$~\kms, respectively, in a model that allows for high-velocity outliers; or $81$, $58$, and $62$~\kms\ for the full population. These velocity dispersions are far smaller than those of the canonical pulsar population, and are similar to older Galactic disk populations.  This suggests that millisecond pulsar velocities are largely attributable to their being an old population rather than being artifacts of their birth and evolution as neutron star binary systems.  The components of these velocity dispersions follow similar proportions to other Galactic populations, suggesting that our results are not biased by selection effects. 

\end{abstract}

\section{Introduction}
\label{sec:intro}
\makeatletter{}Distance and velocity measurements of millisecond pulsars can be used to constrain models of supernova dynamics, binary star evolution, pulsar emission physics, and the ionized interstellar medium.  They can characterize the millisecond pulsar population as a whole, and they can be used to probe the kinematic evolution of the millisecond pulsar population in the Galaxy and its relation to other stellar populations. 

Pulsar timing allows for high precision measurement of millisecond pulsar 
positions, parallaxes, and proper motions.  Pulse times of
arrival (TOAs) measured over the course of a year vary in part due to
the changing time-of-flight of the pulses across the solar system, i.e., the
Roemer delay.  This variation is approximately sinusoidal with a period of
one year.  The phase and amplitude of this pattern can be used to infer the
ecliptic longitude and latitude of the pulsar, respectively.  For
many millisecond pulsars, TOAs can be measured to a precision well
under a microsecond, yielding position measurements with precision of
order milliarcseconds or better.  
Measurements of positions over 
several years can be used to infer proper motions with precision of 
milliarcseconds per year.
Such highly precise TOA 
measurements can also be used to infer pulsar parallaxes, and hence distances,
 out to distances of order a kiloparsec.  
Proper motions
and distances combine to yield two components of the pulsar velocity
vectors.

Canonical (non-millisecond) pulsars are well known to be high velocity objects.
For example, the proper motion study of \cite{Hobbs2005} found that canonical
pulsars have mean speeds of $152\pm 10$~\kms when measured in
one dimension and $246\pm 22$~km/s when measured in two dimensions; from this
they inferred that pulsars are formed in a thin disk with a mean birth speed of $400\pm 40$~km/s.  These high velocities presumably result from asymmetries in
supernova explosions, possibly combined with pre-supernova orbital and space motion of
the pulsar progenitor.

In contrast, early studies of millisecond pulsars showed their space
velocities to be much lower than those of canonical pulsars \citep{Nice1995,
Cordes1997}.  This has been confirmed by, among others, \cite{Hobbs2005},
whose analysis of recycled pulsars (which they defined as having
spin periods $P<100$~ms and spin-down
rates $\dot{P}<10^{-17}$) found their one-dimensional and two-dimensional mean
speeds to be $54\pm 6$~\kms and $87\pm 13$~\kms, respectively. \cite{Gonzalez2011}, whose analysis of millisecond pulsars (which they
defined to have spin periods
$P<10$~ms) found their mean two-dimensional speed to be $108 \pm 15$\,km/s.

The relatively small space velocities of millisecond pulsars are likely
a byproduct of their formation process.  According to the conventional scenario, a millisecond
pulsar is formed as a canonical pulsar and then spun-up to its 
millisecond rotational period via mass accretion from
a binary companion, \citep{Alpar1982}.  If a neutron star in a binary system
is formed with a large kick, the binary is likely to be disrupted, and
the binary accretion needed to form a millisecond pulsar cannot take place.  
Only binary systems with small or fortuitously oriented kicks
survive and allow 
formation of millisecond pulsars.  Alternatively, low-velocity
kicks may result from electron-capture supernovae, and the
millisecond pulsars may be formed via accretion-induced collapse of
a white dwarf during mass transfer in a binary \cite[e.g.,][]{VanDenHeuvel2011,2014FreireTauris}.

One might expect correlations between millisecond pulsar 
kinematics and orbital properties, i.e., different velocity distributions 
for isolated and binary millisecond pulsars; and, among binary millisecond pulsars, a 
dependence of velocity on orbital period or mass function \cite[e.g.,][]{Tauris1996}.
Indeed, \cite{Ng2014} found greater Galactic heights
for lower-mass binary millisecond pulsar systems.  
\cite{Lommen2006} compared isolated and binary
millisecond pulsars and found that they had indistinguishable velocity distributions, but that
observed binary pulsars had a larger Galactic scale height, leading to speculation
that the two populations have different luminosity distributions, with isolated
pulsars weaker and therefore more difficult to detect at large distances; however, \cite{Lorimer2007} argue that this could arise from selection effects in pulsar searches.
\cite{Gonzalez2011}, analyzing a larger sample of millisecond pulsars,
also found the velocity distributions of binary and isolated millisecond pulsars
to be indistinguishable.  

Stellar populations in the solar neighborhood have ellipsoidal
velocity distributions. \cite{Cordes1997} show that analysis of these velocity
distributions can give insight into the dynamical history of millisecond
pulsars, including both the magnitudes of any kicks received at
formation and the diffusion of these pulsars through the Galaxy.
 We adopt Galactocentric coordinates, with
$v_R$, $v_\phi$, and $v_z$ in the direction of the Galactic
center, the direction of Galactic rotation, and the direction
perpendicular to the Galactic disk, respectively,\footnote{These
are often labeled $U$, $V$, and $W$ in the solar neighborhood.
We use alternate notation
to acknowledge that some pulsars in our analysis are far from
the solar neighborhood.}
and we denote their velocity dispersions by $\sigma_R$, $\sigma_\phi$, and
$\sigma_z$.
It is well established theoretically and observationally that,
for any given stellar population, 
$\sigma_R>\sigma_\phi>\sigma_z$, with typical ratios
$\sigma_z/\sigma_R\simeq 0.5$ and $\sigma_\phi/\sigma_R\simeq 0.55\ {\rm to}\ 0.7$
\citep[\S 10.3.2; for simplicity, we ignore vertex deviation]{GalacticAstronomy}.
Analyzing millisecond pulsar velocities
in Galactocentric coordinates has three advantages.
(i) It facilitates comparisons with other stellar populations.
(ii) The consistency of the magnitudes of the different
components can be used to cross-check the measured velocity distributions.  (iii) If 
a component were biased by selection effects---in particular,
pulsar searches are preferentially made along the Galactic plane,
and one could imagine this biasing the distribution of $v_z$---the
other components might still give unbiased results.

The NANOGrav collaboration\footnote{\url{http://www.nanograv.org}} is undertaking
long-term timing of several dozen millisecond pulsars for the purpose of
detecting and characterizing gravitational waves via their perturbation
of millisecond pulsar TOAs on time scales of tens of years (frequencies
of nanohertz).  The detection of gravitational wave signals requires
the modeling and removal of all other phenomena that influence pulse
arrival times, including solar system time-of-flight effects which yield
the astrometric measurements described above.  
The NANOGrav nine-year data release
\citep{Arzoumanian2015}
reports timing observations 
of 37 pulsars collected over nine years, and includes pulse timing
models for all of these pulsars.  In the present paper, we 
analyze the astrometric parameters from that work.

In \S~\ref{sec:obs}, we summarize the observations.  
In \S~\ref{sec:measurements}, we list the parameters resulting from the timing analysis.
In \S~\ref{sec:distances}, we discuss the distances inferred from the timing measurements.
In \S~\ref{sec:mspvelocities}, we combine our newly measured millisecond pulsar
velocities with previous measurements to analyze the kinematics of the millisecond pulsar population as a whole.
In \S~\ref{sec:conclusion}, we summarize our results.

Except where otherwise specified, we define a millisecond pulsar to be a pulsar
with rotational period of 20~ms or less, and a spin-down rate of less than  
$10^{-17}\,{\rm s}\,{\rm s}^{-1}$, a definition that includes all pulsars observed
in the NANOGrav program.  We exclude pulsars in globular clusters
from our study, because no NANOGrav sources are in globular clusters, and because
the formation and dynamics of millisecond pulsars in globular clusters are
different than those of millisecond pulsars in the Galactic disk.

\section{Observations}
\label{sec:obs}
\makeatletter{}\subsection{Overview of the NANOGrav nine-year data set}

We use data from the NANOGrav nine-year data set, which we  briefly summarize 
here.  For full details see \cite{Arzoumanian2015}.  Observations of
37 millisecond pulsars were made using the Arecibo Observatory (AO) and the
Green Bank Telescope (GBT).  The project began with 15 pulsars in 2004, and
new pulsars were added to the project as they were discovered or when
wide-band data acquisition systems enabled sources previously deemed unreliable to be precisely timed.  Pulsars were chosen based on high timing
precision, detectability over a wide frequency range, and expected timing
stability.  The nine-year data set includes observations through 2013. 
Observed time spans of individual pulsars range from 0.6~yr to 9.2~yr.

Each pulsar was observed at approximately monthly intervals.
At every epoch, each pulsar was observed for approximately $25$ minutes each 
using two different radio telescope receivers at widely separated frequencies.  
Such dual-receiver observations
allow for measurement and removal of interstellar and solar-system dispersive
effects.  (In a small number of cases, dual-receiver observations were not
available, but single-receiver observations were made over a wide observing
band.)

Observations in early years of the project used the Astronomical Signal
Processor (ASP) at AO and the Green Bank Astronomical Signal Processor (GASP)
at the GBT; each sampled up to 64~MHz bandwidth, depending on telescope
receiver capability.  In the later years, observations used the Puerto Rican
Ultimate Pulsar Processor (PUPPI) at AO and the Green Bank Ultimate Pulsar
Processor (GUPPI) at the GBT; each provided up to 800~MHz bandwidth, again
depending on telescope receiver capability.

The data were polarization-calibrated, cleaned of radio frequency interference, 
and analyzed to produce TOAs using a standardized pipeline\footnote{\url{
http://github.com/demorest/nanopipe}} that made extensive use of the
{\sc psrchive} software package
\citep{psrchive}\footnote{\url{http://psrchive.sourceforge.net}}. 
The sets of TOAs were fit to timing models using the
{\sc tempo} and {\sc tempo2} packages.  We found that these packages yielded
essentially identical results, and we used {\sc tempo} for the work in the
present paper \footnote{\url{http://tempo.sourceforge.net}}.

The timing models contain standard spin-down, astrometric, and (where
appropriate) binary models.  The astrometric parameter measurements are the
basis for the present paper and are described in more detail in
\S\ref{sec:measurements}.  Criteria for inclusion of specific parameters
in the timing model for each pulsar are given in \cite{Arzoumanian2015}.

In the timing analysis, Earth motion around the solar-system barycenter was
modeled using the JPL DE421 planetary ephemeris \citep{DE421}.  
TOAs were measured using hydrogen-maser clocks at the observatories
and were transformed to Universal Time and then to Barycentric Dynamical Time using standard
techniques.

Separate values of dispersion measure were fit independently at every
observing epoch, with a few exceptions (see \cite{Arzoumanian2015} for
details).  This results in timing parameter uncertainties that are larger than
would be found using smoothed dispersion measure models, but is necessary to
avoid contamination of our results by variations in interstellar and solar
wind dispersion.  The latter, in particular, is not easily modeled, and is
highly covariant with the astrometric parameters of interest to this paper.

A novel noise model was used to account for any aspects of the timing data
that did not fit the standard timing model within the expected measurement
uncertainties.  The noise model
incorporated white noise, correlated noise in simultaneously collected
TOAs, and red noise (as needed for a few sources).  Details are given in
\cite{Arzoumanian2015}.

The full data set, including TOAs and timing parameters of all pulsars, is
available online\footnote{\url{http://data.nanograv.org}}.

\subsection{Data set modifications for this work}

We made a small number of modifications to the NANOGrav nine-year data
set for the present work.

The data sets of PSRs J1741+1351, J1853+1303, J1910+1256, J1944+0907, and
B1953+29 contain lengthy spans in which observations were made using only a
single, narrow-band receiver, eliminating the possibility of monitoring
and correcting for
dispersion measure variation over those spans.  We excised the TOA
measurements from those data spans.   We re-evaluated the timing models for
these pulsars using the remaining TOAs, following the same guidelines used by
\cite{Arzoumanian2015} for noise models, parameter inclusion, etc.

The timing solution for PSR J2317+1439 in the NANOGrav nine-year data release
included secular variations of the Laplace-Lagrange orbital elements, which
are formally significant in the fit, but which are physically implausible.  We
removed those parameters from the timing solution.  We then re-evaluated the
timing model using the guidelines of \cite{Arzoumanian2015}, following which
we added Shapiro delay parameters (now significant) and eliminated the secular
variation in the projected orbital semi-major axis (now not significant).
Details of the analysis of this and other binaries will be given in a
forthcoming paper (Fonseca et al., in preparation).

The modified timing solutions used for this paper are bundled with the 
NANOGrav nine-year data release files\footnote{\url{http://data.nanograv.org}}.

\section{Measured astrometric parameters}
\label{sec:measurements}
\makeatletter{}\subsection{Positions}

The timing analysis of the NANOGrav nine-year data set parameterizes pulsar
positions in ecliptic coordinates.  These are natural coordinates for pulsar
timing astrometry, as ecliptic longitude and latitude are nearly orthogonal parameters in
the timing fit.  Table~\ref{tab:equatorial} lists the ecliptic and equatorial positions of
all of the pulsars in the data set.  The epoch of each position was chosen
near the middle of its data set in order to minimize covariance between
position and proper motion parameters.  The reference frame for these position
measurements is the reference frame of JPL planetary ephemeris DE421, which is
oriented to the International Celestial Reference Frame \citep{DE421}.
The ephemeris coordinates were transformed into ecliptic coordinates
by a rotation of $23^\circ 26\arcmin 21\farcs 406$, the IERS2010 obliquity of the ecliptic.

\begin{figure*}[t]
  \centering
    \begin{minipage}{.3\textwidth}
      \includegraphics[width=\linewidth]{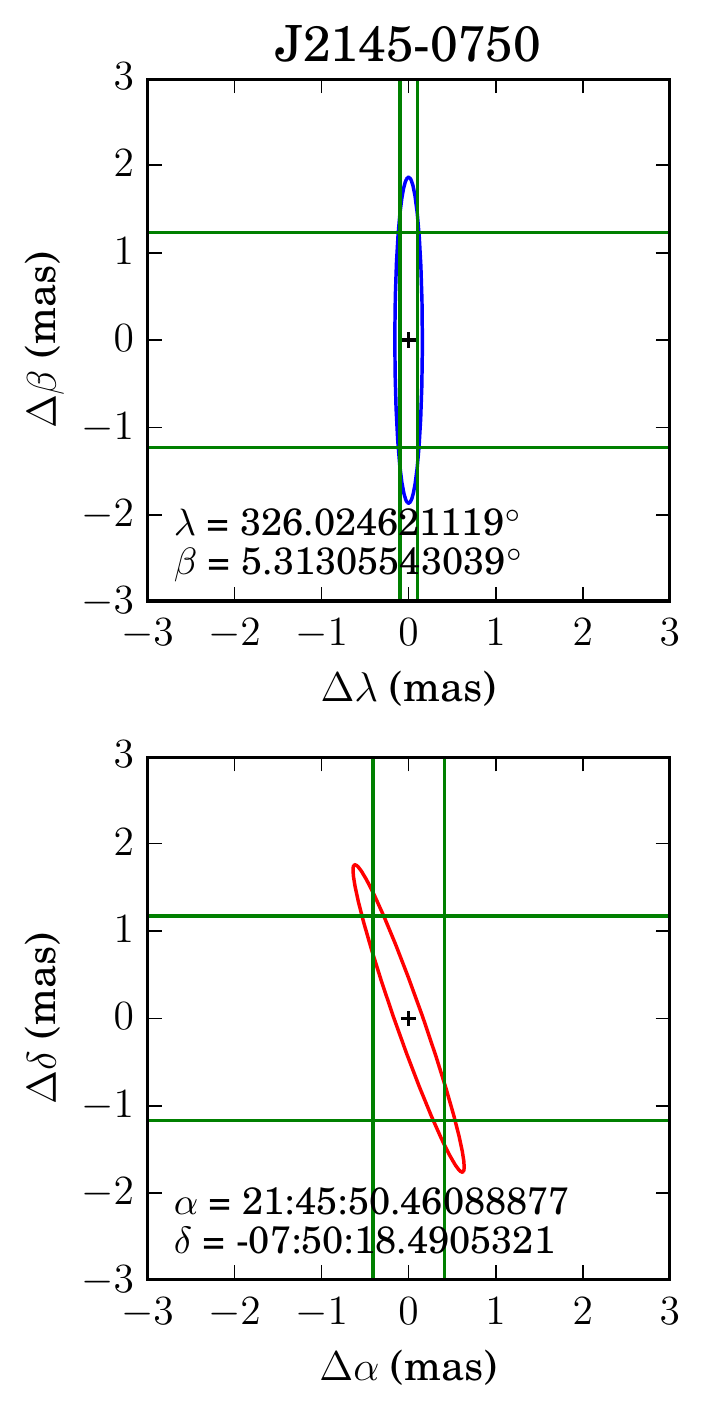}
  \end{minipage}  \hspace{.05\linewidth}
  \begin{minipage}{0.315\textwidth}
      \includegraphics[width=\linewidth]{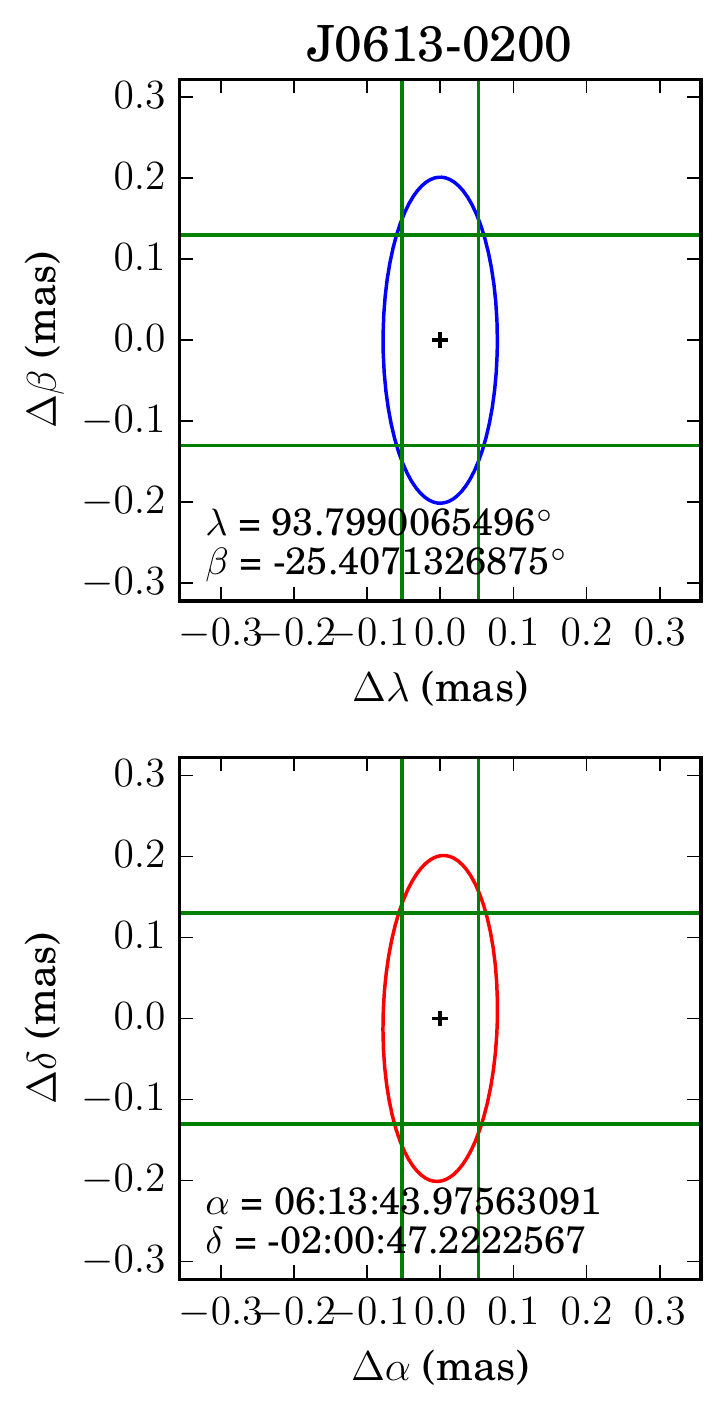}
  \end{minipage}
  \caption{Examples of $1\sigma$ error ellipses for pulsar position in equatorial coordinates, shown in red, and ecliptic coordinates, shown in blue. Green lines indicate $1\sigma$ uncertainties in mas, and hence they show the error region that can be inferred when the coordinates are reported in each coordinate system.  For J2145-0750, $\sigma_{\lambda} = 0.10$ mas, $\sigma_{\beta} = 1.23$ mas, $\sigma_{\alpha} = 0.41$ mas, and $\sigma_{\delta} = 1.17$ mas.  The error region is much larger in equatorial coordinates due to the covariance between right ascension and declination.   For J0613-0200: $\sigma_{\lambda} = \sigma_{\alpha} = 0.052$ mas, and $\sigma_{\beta} = \sigma_{\delta} = 0.13$ mas, and the error regions are similar in size in the two coordinate systems. The severity of the covariance in equatorial coordinates, and hence the enlargement of the error ellipse when expressed in equatorial coordinates, depends on the pulsar's ecliptic latitude and its proximity to the equinoxes.}
  \label{fig:position_covariance}
\end{figure*}

For observational convenience, Table~\ref{tab:equatorial} also lists the
positions in equatorial coordinates.  The equatorial positions were determined
by fitting timing solutions without rotating the planetary ephemeris into
ecliptic coordinates.  As shown in Figure~\ref{fig:position_covariance},
uncertainties in equatorial coordinates are generally larger than
uncertainties in ecliptic coordinates.
This is particularly true when one ecliptic coordinate is
measured much more precisely than the other (e.g., for pulsars along the
ecliptic).

\subsection{Proper Motions}

Proper motion, $\mu_{\alpha} = \dot{\alpha}~{\rm cos}\delta$ and $\mu_{\delta}
=\dot{\delta}$, was included in the timing model for 35 of the 37 pulsars in our
data set.  The two pulsars for which proper motion was excluded have less than
one year of timing measurements in our data set, making it impossible to
measure their proper motions.  Lengths of all of the data sets are given in
Table~\ref{tab:propmo}.

The proper motions from timing analyses in both ecliptic and equatorial
coordinates are listed in Table~\ref{tab:propmo}.  As with the position
measurements, ecliptic coordinates provide the best separation of proper
motion into orthogonal components, and hence the smallest uncertainties. The
table also includes proper motions in Galactic coordinates.

The best previous proper motion measurements in ecliptic coordinates are
listed in Table~\ref{tab:propmo}.  All the new and previous measurements agree
to within $3\sigma$ except for two pulsars (PSRs~J1909$-$3744
and J2145$-$0750) which have discrepancies between $3\sigma$ and $4\sigma$
for at least one component of proper motion.  The proper motions of these
two pulsars are known to high precision, and these small discrepancies have
little practical impact.  We speculate that the differences may be due to
differences in dispersion measure variation models (as we discuss for parallax
measurements in \S~\ref{sec:pxcomparison}), or differences in solar system
ephemerides used in the timing analysis.

\begin{figure*}
  \centering
  \includegraphics[scale=0.45]{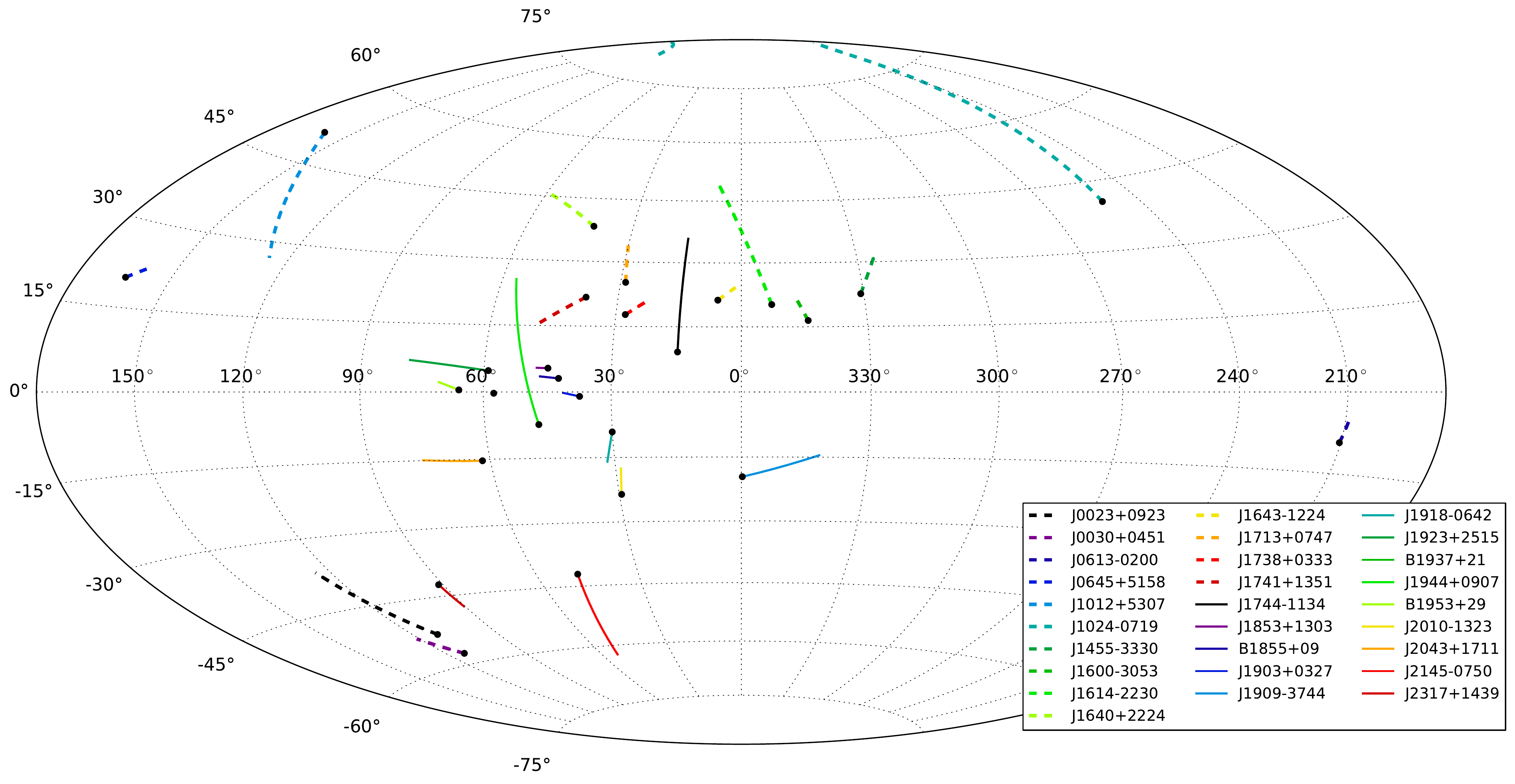}
  \caption{Path of pulsar Galactic motion from 5 Myr ago to present, shown as black dots. Millisecond pulsars with greater than 5-sigma significant proper motion measured by NANOGrav are shown.  Color and line style (solid or dashed) are used to identify individual pulsars but have no other significance.}
  \label{fig:Galactic_velocity}
\end{figure*}

Our measured proper motions are illustrated in Figure~\ref{fig:Galactic_velocity}.
We discuss the derivation of two- and three-dimensional space velocities from these
proper motions in \S~\ref{sec:mspvelocities}.

\subsection{Parallaxes}

\begin{figure*}
  \centering
  \begin{minipage}{\textwidth}
    \centerline{\includegraphics[scale=0.8]{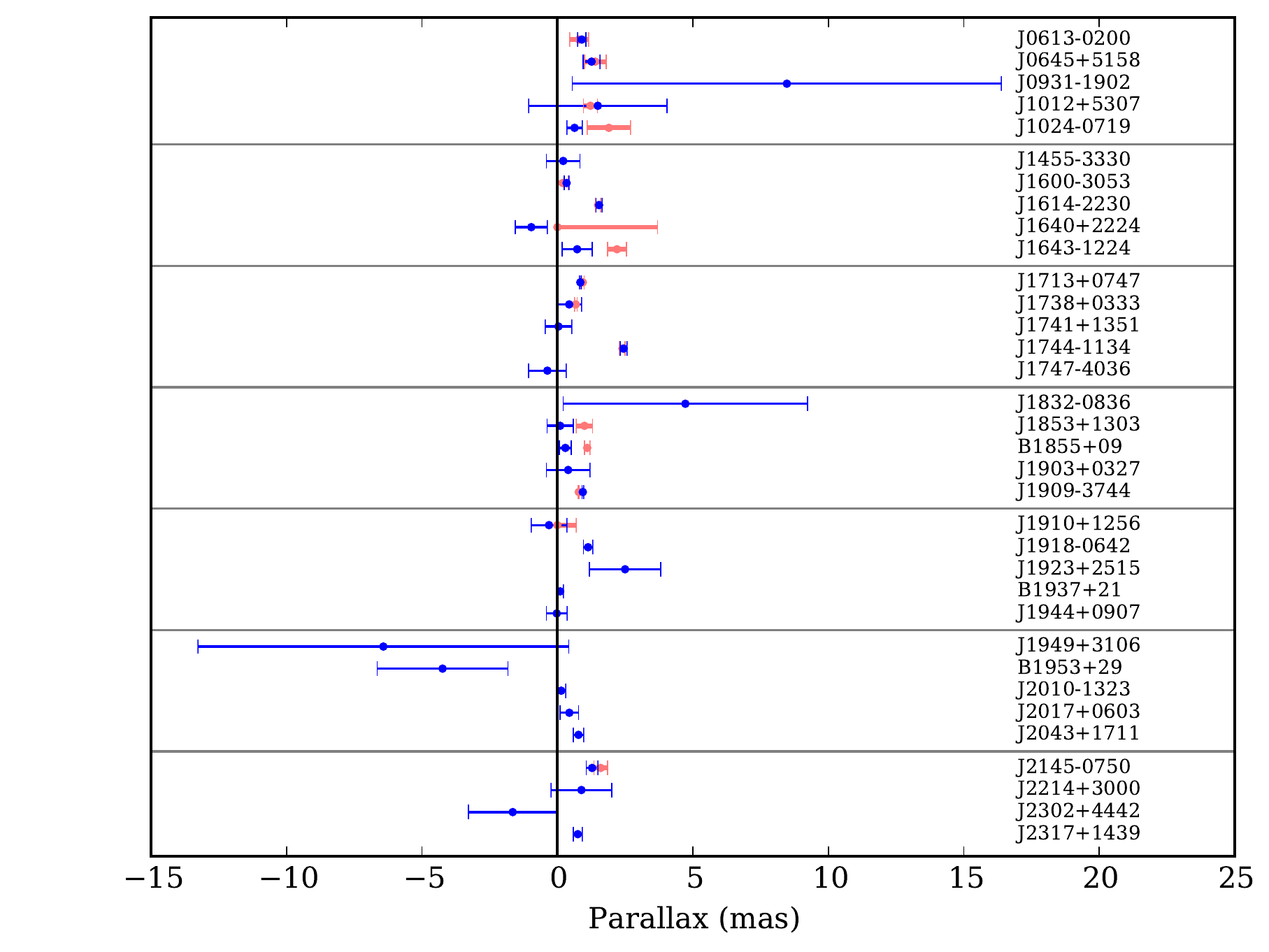}}
  \end{minipage}
  \begin{minipage}{\textwidth}
    \centerline{\includegraphics[scale=0.8]{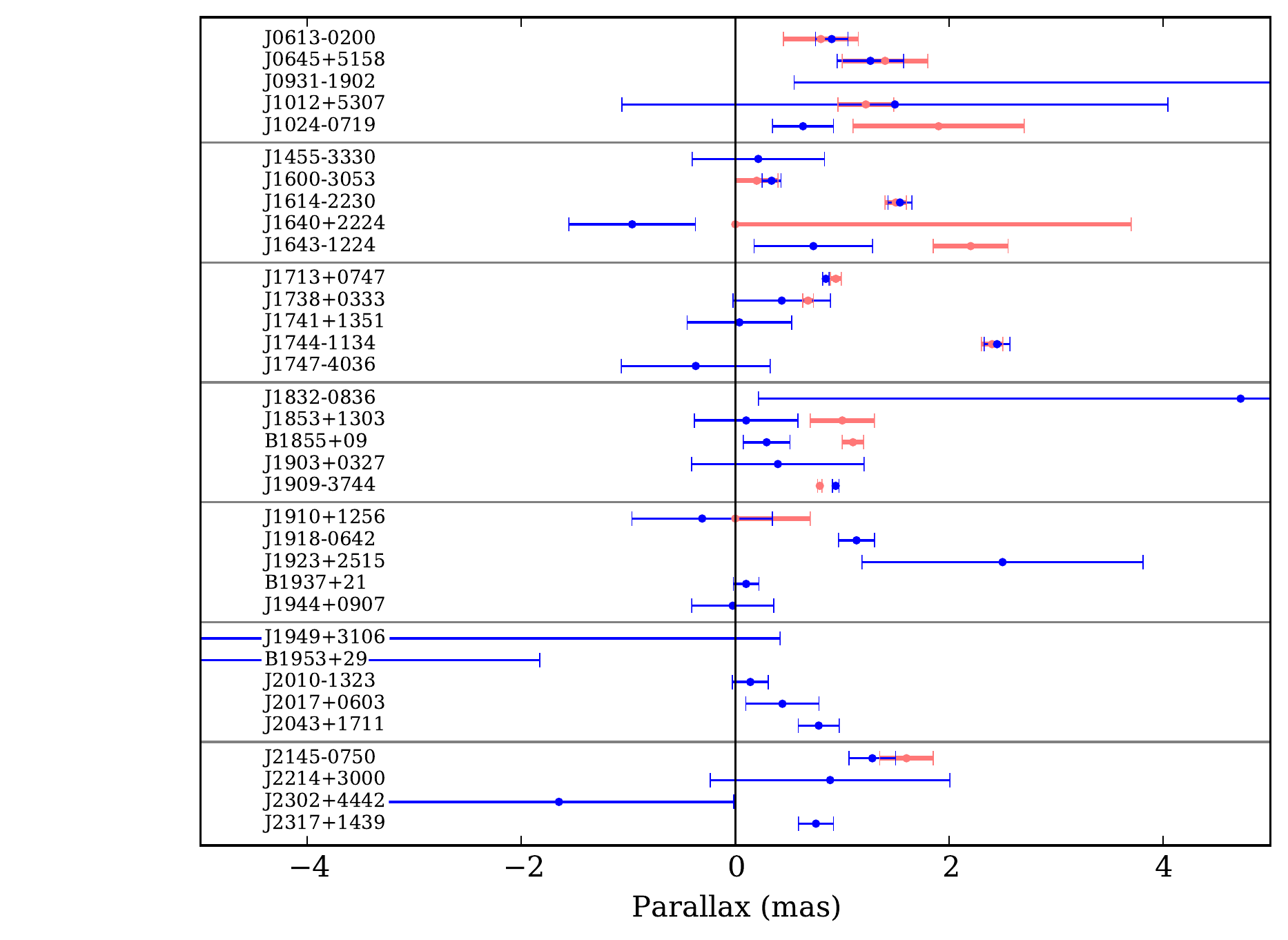}}
  \end{minipage}
  \caption{{\bf Top:} NANOGrav nine-year parallax values shown in blue in comparison to the previous best measurement or limit shown in faded red. {\bf Bottom:} Top plot restricted to narrower parallax range.
  \label{fig:px}
  }
\end{figure*}

We included parallax, $\varpi$, as a free parameter in the timing model for 
each pulsar,
whether or not the measurement was statistically significant.
The timing analysis allowed both negative and positive parallaxes.  Although
negative parallaxes are non-physical, allowing them in the timing solutions
provides a useful
a check on the reliability of the measurements (\S\ref{sec:pxlowsignificance}).
The measured parallax values are listed in
Table~\ref{tab:px} and shown in Figure~\ref{fig:px}.  Previous measurements
for these pulsars are also included in the Table and Figure.

There were no previously reported parallax measurements or limits for 20
of the 37 pulsars.  Of these 20 sources, two provided significant
new parallax measurements: PSR~J1918$-$0642, $\varpi=1.1\pm 0.2$\,mas; and
PSR~J2043+1711, $\varpi=0.8\pm 0.2$\,mas. 

\subsubsection{Comparison with previous measurements}
\label{sec:pxcomparison}

Of the 17 pulsars with previous measurements or limits, we improved the precision of four measurements by a factor of two or better: PSRs J0030+0451, J0613$-$0200, J1600$-$3053, and J1614$-$2230.

For PSR~J1713+0747, in addition to the previous timing parallax value of 0.94(5) given in
Table~\ref{tab:px}, there are two other previous measurements of interest.
One is from an analysis of 21 years of data from this pulsar by
\cite{Zhu2015}, which obtained a parallax of $\varpi=0.85\pm0.03$ mas, identical
to our own measurement.\footnote{\cite{Zhu2015} also obtained a slightly
different value in an analysis using the TEMPO2 software package, with which
they employed a different binary model for this particular pulsar.}  This is
not surprising, as the vast majority of the TOAs in the \cite{Zhu2015}
analysis were from the same NANOGrav nine-year data set that is being used for
the present work.  The other previous parallax measurement of J~1713+0747
is from the interferometric VLBA measurements of \cite{Chatterjee2009}, which
obtained $\varpi=0.95\pm0.06$ mas, in reasonable agreement ($1.5\sigma$) with our
value.

In most other cases, our parallax values also agree with previous measurements.
However, 
there are discrepancies of 2$\sigma$ or more (i.e., disagreement at 95\%
confidence) between our measurements and the previous measurements for three
sources: PSRs J1643$-$1224, B1855+09, and J1909$-$3744.  A possible
explanation for these discrepancies is the difference in treatment of
dispersion measure (DM) variations.  Small variations in the ionized component of
the interstellar medium and the solar wind along the pulsar--Earth
line-of-sight significantly affect TOAs.  To minimize
contamination of the timing analysis from DM variations on the half-yearly 
time scale of the parallax signal (as well as other time scales of interest
to pulsar modeling),
we fit for DM at every epoch of observation.  Since the DM
variations are covariant with the parallax signal in the timing fit, this 
increases formal uncertainties on our parallax measurements,
but it also makes the measurements more robust.

The previous parallax analysis of two of the discrepant
sources---PSRs J1643$-$1224, and B1855+09---assumed a constant value of
DM over the entire data set (see references in Table~\ref{tab:px}).  The
third, PSR~J1909$-$3744 \citep{Verbiest2009}, used the smoothed DM time series
model of \cite{You2007}, which includes a smoothing time scale longer
than the half-year period of the parallax signature in the timing model.  To
test the effect of using constant or smoothed DM variation models, we ran trials of timing analyses of
these sources fitting (i) a constant DM and (ii) a model of DM as a up-to-fifth-order 
polynomial in time over the entire data sets.   These tests yielded
changes in measured parallax values by amounts ranging from
$0.2\sigma$ to $10\sigma$, with three of the four values moving by at least
$1\sigma$.  We therefore believe that limited DM variation modeling is a
plausible explanation for the discrepancies between previously measured 
parallax values and our measurements.

\subsubsection{Measurements with low significance}
\label{sec:pxlowsignificance}

\begin{figure*}
\centering
\includegraphics[scale=0.8]{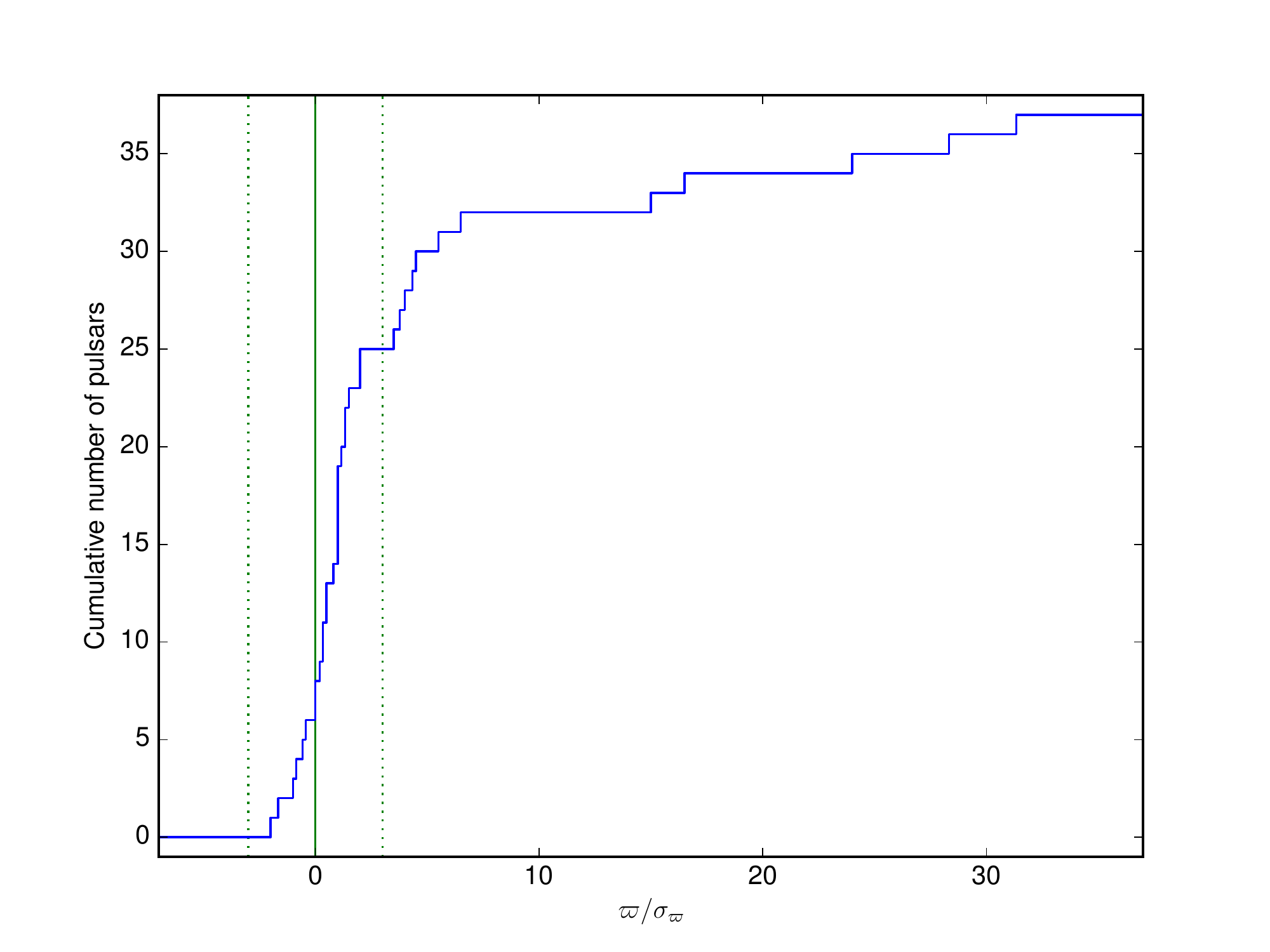}
\caption{Cumulative number of pulsars as a function of parallax measurement significance,
$\varpi/\sigma_\varpi$.  The dotted green lines delimit $\pm 3\sigma$.
\label{fig:px_cdf}
}
\end{figure*}

The timing analyses of a majority of the pulsars we analyzed did not yield significant parallax 
detections.  To illustrate the nature of these non-detections, we sorted the pulsars
by parallax measurement significance, $\varpi/\sigma_\varpi$, and plotted the cumulative
distribution of these values in Figure~\ref{fig:px_cdf}. Of the 37
parallax measurements, 12 are significant positive detections of parallax,
$\varpi/\sigma_\varpi > 3$.  No sources had significant negative detections of
parallax, $\varpi/\sigma_\varpi < -3$.  
(Although negative parallax values are unphysical, corresponding to pulse wavefronts 
with concave rather than convex curvature relative to the pulsar, they were
allowed in the timing analysis.  A significant negative parallax measurement 
would indicate a problem with the data or the timing model.) 
Of the 25 pulsars
with non-detections, only 7 measurements are negative, all but two within
$1\sigma$ of zero, and the remaining 18 are positive.  This bias toward
positive values indicates that most of these are indeed real, physical
measurements, and significant parallax values should be attainable
for many of them with only moderately larger data sets.

\section{Distances}
\label{sec:distances}
\makeatletter{}\subsection{Distances from parallax measurements}

We now use the parallaxes in Table~\ref{tab:px} to analyze distances to 30 of
the 37 pulsars.  The timing data sets of the other 7 pulsars span less than 2
years, which is typically not sufficient time to disentangle position, proper
motion, and parallax in the timing analysis.  Indeed, none of these sources
had nominally significant parallax measurements, and 3 of them had negative
parallaxes.

For the 12 pulsars with significant parallax measurements,
$\varpi/\sigma_\varpi>3$, we calculated distances as listed in the upper
portion of Table~\ref{tab:pxlimits}.  For each of these pulsars, the central
distance value, upper limit, and lower limit given in the table were
calculated using $d=\varpi^{-1}$, where $\varpi$ was the 84\%, 50\%, and 16\%
points in the measured parallax distribution, respectively, corresponding
to the 16\%, 50\% and 84\% points in the distance distribution.

For the 18 remaining pulsars, we used 95\% confidence upper limits on parallax
to compute 95\% confidence lower limits on distance.  To find the upper limits
on parallax, we first verified that changes in assumed parallax values in the
timing solution yielded changes in goodness-of-fit $\chi^2$ values appropriate 
for parallax values drawn from a Gaussian distribution.  We then found the 95\% confidence upper
limits on $\varpi$ by calculating the value of $\varpi$ at which 95\% of the
area under the Gaussian distribution of the parallax was at lower values, but
restricting the integral to positive values of parallax.  The resulting parallax
limits
are listed in Table \ref{tab:pxlimits}. The corresponding $95\%$ confidence
lower limit on distance for each source was then calculated to be the
reciprocal of this value. These are also listed in Table \ref{tab:pxlimits}.

The Lutz-Kelker bias \citep{LutzKelker1973} may be significant for the distance
measurements of pulsars with large uncertainties on parallax. The bias can be
incorporated as a correction to parallax and distance values through methods
outlined by \cite{Verbiest2010} and \cite{Verbiest2012}. 
The Bayesian analysis used under these methods assumes a uniform prior
distribution of a pulsar's position throughout three-dimensional space.  Since
there is more three-dimensional space at larger distances than at smaller
distances, this has the effect of moving the posterior distribution
of a pulsar's distance to values larger than that calculated from
$d=\varpi^{-1}$.  This is problematic for pulsars for which 
there are only upper limits on parallax, as this allows a finite probability 
of $\varpi_{\rm measured}=0$, which corresponds to infinitely large distances, at
which there is infinite parameter space.  To achieve a tractable result,
some additional prior must be invoked to cut off the likelihood of the
pulsar being at a large distance.  One possibility is to use a model of
the Galactic distribution of millisecond pulsars, but
we want our results to stand independent of {\it a priori}
beliefs about the Galactic distribution of millisecond
pulsars; further, we found that attempts to use prior Galactic distribution
models tended to give results that were dominated by the details of the
model.  Another possibility is to model the sensitivity of pulsar  
search programs, which find nearby pulsars more readily than distant
pulsars; however, we found that use of such a prior gave results
that were highly dependent on the
choice of millisecond pulsar luminosity distribution, which is
not well known.

For these reasons, we elected not to apply any corrections to the distances
in Table~\ref{tab:pxlimits}.  For sources for which we give lower limits
on distances (lower portion of Table~\ref{tab:pxlimits}), this is a conservative
approach from a Lutz-Kelker perspective, in the sense that the Lutz-Kelker
bias would place these pulsars still further than our lower-limit distances.
For the twelve sources for which we give distance measurements (upper portion
of Table~\ref{tab:pxlimits}), we tested the effect of Lutz-Kelker corrections
by running the measurements through the Lutz-Kelker bias code from
\cite{Verbiest2012}\footnote{\url{http://psrpop.phys.wvu.edu/LKbias/}}.  We
found the differences in calculated distances to be small; for example, six of
the central distance measurement values were unchanged, and the other six all
changed by less than $1\sigma$ (typically much less).

\subsection{Distances from orbital period derivatives}
\label{sec:pbdotdistances}

\newcommand{\pbdot}{\dot{P}_{\rm b}} 
\newcommand{\pb}{{P}_{\rm b}}
\newcommand{\pdot}{\dot{P}}

In addition to timing parallax, there are other means by which distances
can be inferred from pulsar timing.  One such approach uses the observed
properties of binary pulsars.  
The observed orbital period, $P_{\rm b}$, 
of a binary pulsar undergoes a time change $\dot{P}_{\rm b}$ due to mass loss from the system (negligible here), 
general relativistic phenomena, and 
kinematic effects \citep{dt91}, 
\begin{equation}
    \left(\frac{\pbdot}{\pb}\right)_{\rm obs} 
  = \left(\frac{\pbdot}{\pb}\right)_{\rm GR} 
  + \left(\frac{\pbdot}{\pb}\right)_{\rm kin}. 
\label{eqn:pbdot1}
\end{equation}
The relativistic term, $\left(\dot{P}_b/P_b\right)_{\rm GR}$, is the orbital
decay due to emission of gravitational radiation, which depends on the
component masses and other orbital elements.  The kinematic term,
$\left(\dot{P}_b/P_b\right)_{\rm kin}$, is due to the relative acceleration of
the binary system and the Sun, which causes changes in the Doppler shift of
the observed period.  The relevant expression from \cite{dt91}, here modified
to apply to pulsars off the Galactic plane \citep{Nice1995}, is
\begin{equation}
  \left(\frac{\pbdot}{\pb}\right)_{\rm kin} 
    =  \frac{a_z(z)}{c} - \cos b \left(\frac{\Theta_0^2}{cR_0}\right)\left(\cos l + \frac{\beta}{\sin^2l+\beta^2}\right)+\frac{\mu^2d}{c},
\label{eqn:pbdot2}
\end{equation}
where the terms on the right hand side are as follows. The first term 
is the acceleration of the binary system in the Galactic potential perpendicular
to the disk, $a_z$, which depends on the perpendicular distance $z = d\sin b$
, where $b$ is the Galactic latitude, and $c$ is the speed of light.
The second term is the line-of-sight
acceleration due to differential rotation in the Galaxy; here $R_0$ is the
Galactocentric distance of the solar system; $\Theta_0$ is the circular
rotation speed of the Galaxy at $R_0$; $l$ is the Galactic longitude; and
$\beta = (d/R_0)\cos b - \cos l$. 
The third term is the
apparent acceleration due to proper motion \citep{shk70}.  Each of these terms depends
on the distance to the pulsar; as \cite{bb96} pointed out, given a measurement
$(\pbdot/\pb)_{\rm obs}$ and a calculated $(\pbdot/\pb)_{\rm GR}$, one can
solve equations~\ref{eqn:pbdot1} and ~\ref{eqn:pbdot2} to infer the distance
to the pulsar.  This method has been used to compute high-precision distances
to PSRs J0437-4715 \citep{vbs+08} and B1534+12 \citep{fst14}.  In cases where
the measurement of $\dot{P}_{\rm b}$ is marginal or not significant, this method
can be used to place an upper limit on the pulsar distance.

There are 25 binary pulsars in the NANOGrav nine-year data set.  A full
analysis of these systems will be presented in Fonseca et al.~(in
preparation).  For the present work, we considered 22 of these pulsars which
have been observed for at least two years.  For each of these, we (i) fit for
$\pbdot$ in the timing solution of the pulsar, (ii) calculated the expected
$\pbdot$ as a function of distance according to equations~\ref{eqn:pbdot1}
and~\ref{eqn:pbdot2}, (iii) compared the calculated $\pbdot$ values with the
measured value and its uncertainty to find a probability associated with that distance, and (iv)
used the probabilities to calculate either the distance and its uncertainty or a
95\% upper limit on the distance.  For the $\pbdot$ calculations, the GR term
is negligible for wide binary systems; for tight binaries, we
estimated $(\pbdot)_{\rm GR}$ using masses from analysis of the Shapiro delay
in each system (Fonseca et al., in prep), except for the case of PSR~J1012+5307,
where we used masses derived from the optical measurements of \cite{Callanan1998}.  For the acceleration
toward the Galactic disk, $a_z(z)$, we used the model of the Galactic
potential given in equation 17 of \cite{Lazaridis2009}, based on the Galactic
potential of \cite{Holmberg2004}; other Galactic potential models, such as
that of \cite{kg89}, give essentially identical results.  
We used $R_0=8.34$~kpc and
$\Theta_0$=240~\kms\ \citep{Reid2014}. 
For the proper
motion term in the $\pbdot$ calculations, we used the proper motions given in
Table~\ref{tab:propmo}.

We found that 15 of the 22 pulsars had distance upper limits greater than 
15~kpc.  Since such limits are of little interest, and since
this is beyond any reasonable extrapolation of the Galactic rotation
and acceleration models, we did not analyze these further.
Of the seven remaining pulsars, we measured significant distance constraints
(at least $2\sigma$) for two sources, and we placed distance upper limits on
the other five sources ($d_{\dot{P}_b}$ in Table~\ref{tab:pbdotdistances}.)  In all cases, these
measurements are consistent with the distance measurements or limits we found
via parallax measurement (Table~\ref{tab:pxlimits}).

The most precise $\pbdot$ distance measurement is that of PSR~J1909$-$3744, 
$1.11\pm0.02$~kpc. This is a good match for our parallax distance
of $1.07\pm 0.03$~kpc.  The best previous parallax measurement,
$\varpi=0.79\pm 0.04$~mas \citep[$2\sigma$,][]{Verbiest2009}
gave a somewhat larger distance of $1.26\pm 0.03$~kpc \citep[$1\sigma$,][]{Verbiest2012}.
As discussed above, we suggest the difference between previous and presently
measured parallaxes may result from differences in the dispersion measure
model used in the timing analysis.  The distance derived from $\pbdot$
is relatively impervious to changes in the dispersion measure model, 
so it provides a good verification of our parallax distance measurement.

Finally, we note that $\pbdot$ values of two sources, PSRs J1012+5307 and 
J1713+0747, have been measured more precisely than in the present paper  \citep{Lazaridis2009, Zhu2015}.
Those works focused on the use of $\pbdot$
measurements to test relativistic gravity rather than measuring distances.
Using those $\pbdot$ measurements in our distance algorithms above gives
distance upper limits of $d<4.8$~kpc and $d<17$~kpc, for PSRs~J1012+5307 
and J1713+0747, respectively.
These constraints, while weak, are consistent with our parallax distance measurements and limits.

\subsection{Distance constraints from rotation period derivatives}
\label{sec:pdotdistances}

Another set of constraints on pulsar distances arises from their spin-down
rates \citep{Nice1995}.  A variant on equation~\ref{eqn:pbdot1} can be applied
to a pulsar rotation period, $P$, and spin-down rate, $\dot{P}$.  The observed
spin-down rate, $\dot{P}_{\rm obs}$ depends on both the intrinsic spin-down
rate, $\dot{P}_{\rm int}$, and kinematic corrections:
\begin{equation}
    \left(\frac{\pdot}{P}\right)_{\rm obs} 
  = \left(\frac{\pdot}{P}\right)_{\rm int} 
  + \left(\frac{\pdot}{P}\right)_{\rm kin}. 
\label{eqn:pdot1}
\end{equation}
The kinematic term, $(\pdot/P)_{\rm kin}$, obeys the same
expression as equation~\ref{eqn:pbdot2}, substituting rotational for orbital parameters,
\begin{equation}
  \left(\frac{\pdot}{P}\right)_{\rm kin} 
    =  \frac{a_z(z)}{c} - \cos b \left(\frac{\Theta_0^2}{cR_0}\right)\left(\cos l + \frac{\beta}{\sin^2l+\beta^2}\right)+\frac{\mu^2d}{c}.
\label{eqn:pdot2}
\end{equation}
If we presume that millisecond pulsars are powered by rotational energy loss,
then $(\pdot/P)_{\rm int}>0$.  From this, equation~\ref{eqn:pdot1} implies
$(\pdot/P)_{\rm kin} < (\pdot/P)_{\rm obs}$, so the observed spin parameters
place an upper limit on this kinematic term.  The right hand side of
equation~\ref{eqn:pdot2} is generally a monotonically increasing function of
distance, so the upper limit on  $(\pdot/P)_{\rm kin}$ sets an upper limit on
distance.

We used equations~\ref{eqn:pdot1} and~\ref{eqn:pdot2} to place upper limits on
distances to the 30 pulsars in the NANOGrav nine-year data set that have been
observed for at least 2 years, so that they have well-measured proper motions.
We used the Galactic rotation and acceleration parameters described in
\S\ref{sec:pbdotdistances}.  We found that 12 of these pulsars had distance
limits greater than 15~kpc; we did not pursue these further.  The spin-down
distance limits, $d_{\dot{P}}$, of the remaining 17 pulsars are listed in
Table~\ref{tab:pdotdistances}.\footnote{For PSR~J1909$-$3744,
in addition to the distance range allowed by the entry in the Table~\ref{tab:pdotdistances},
0 to 1.375~kpc, there are additional allowed solutions at $\sim 9$~kpc,
where the flat Galactic rotation curve model implies
very large accelerations.  Since this pulsar is well established to
be much closer than 9~kpc, we do not include this solution in the table.}

Most of the spin-down distance constraints agree with the distances derived
from parallax measurements.  Indeed, all of the parallax distances measured
with at least 3$\sigma$ precision (Table~\ref{tab:pxlimits}) fall within the
spin-down distance limit constraints (Table~\ref{tab:pdotdistances}).
The pulsar for which the spin-down distance limit comes closest to the
parallax distance is PSR~J1909$-3744$, for which the spin-down limit is
$d_{\dot{P}}<1.375$~kpc and the parallax distance is $d_{\varpi}=1.07\pm
0.03$~kpc, providing a good check on the parallax distance.

For the pulsars for which we have only an upper limit on parallax, and
therefore a lower limit on distance (Table~\ref{tab:pxlimits}), all but
one are in agreement with the spin-down distance limits.  In these cases,
the pair of distance constraints bracket the actual distance to the pulsar.

For PSR~J1024$-$0719, the spin-down distance upper limit, $d_{\dot{P}}<0.427$~kpc, disagrees
with the parallax distance lower limit, $d_{\varpi}>0.91$~kpc.  This is perplexing.
The spin-down distance limit should be robust, presuming that the pulsar
is losing energy and not subject to external acceleration.  We have 
run several tests on our parallax measurement, and we consistently
find low parallax upper limits (i.e., large distance lower limits).
We discuss this source further in Appendix~\ref{app:J1024}.

\subsection{Distances and Galactic electron density}
\label{sec:ne2001}

\begin{figure}
  \centering
  \includegraphics[scale=0.6]{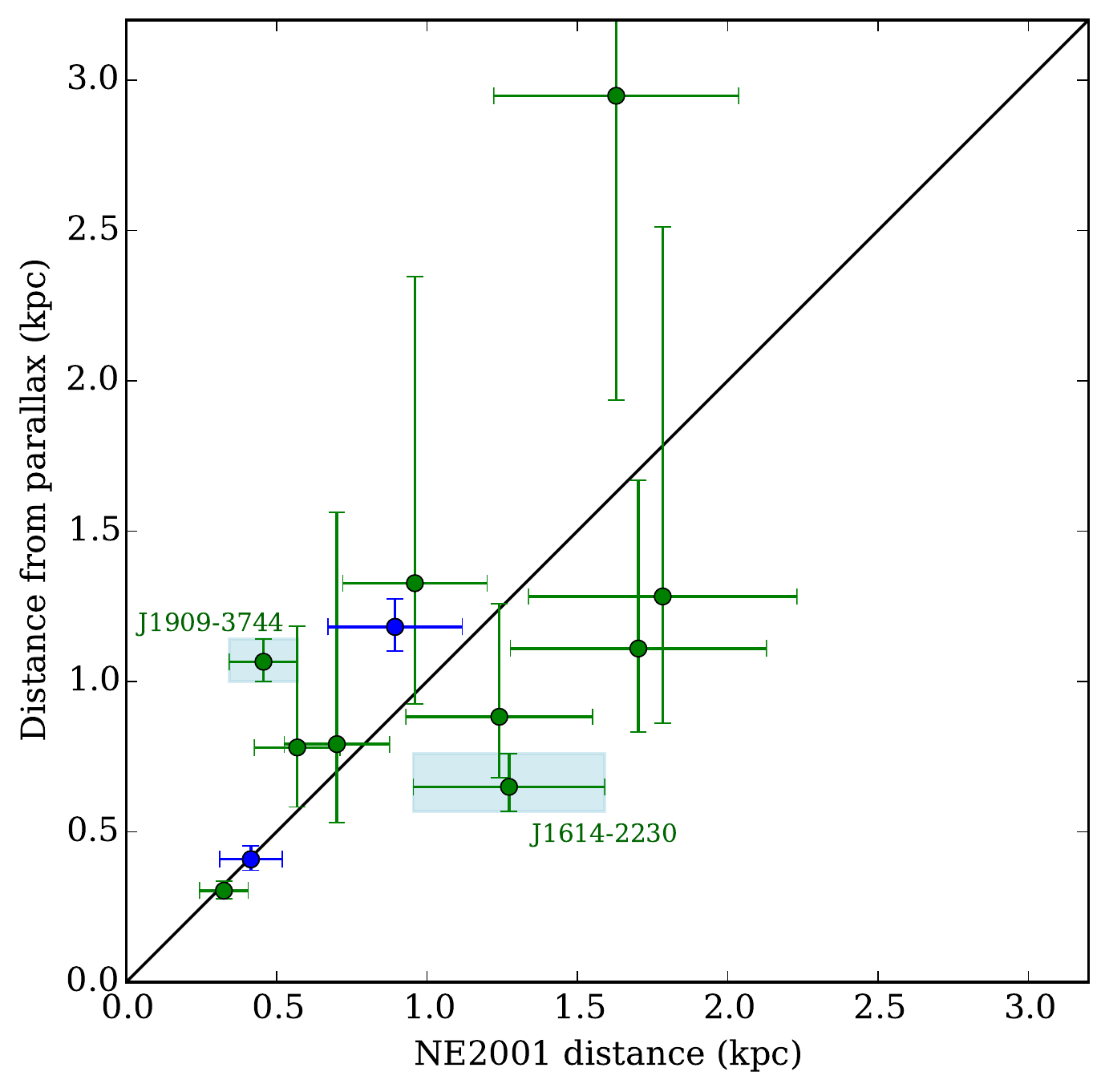}
  \caption{Distances from parallax measurements versus distances from the NE2001 dispersion measure model for pulsars with significant parallax detections in our data set (upper part of Table~\ref{tab:pxlimits}).  Parallax distances are plotted as $2\sigma$ uncertainties; NE2001 dispersion model distances are plotted with 25\% uncertainties.  Two pulsars for which there is no distance that is both within $2\sigma$ of the parallax distance and within 25\% of the NE2001 distance are indicated by name and by light blue uncertainty boxes. Previous parallax measurements of PSRs~J1713+0747 and J1744-1134 were used as input data to the NE2001 model; our measurements of these pulsars are shown in blue.  All of our other measurements are shown in green.
  \label{fig:sigpxvsdm}
}
\end{figure}

\begin{figure}
  \centering
  \includegraphics[scale=0.6]{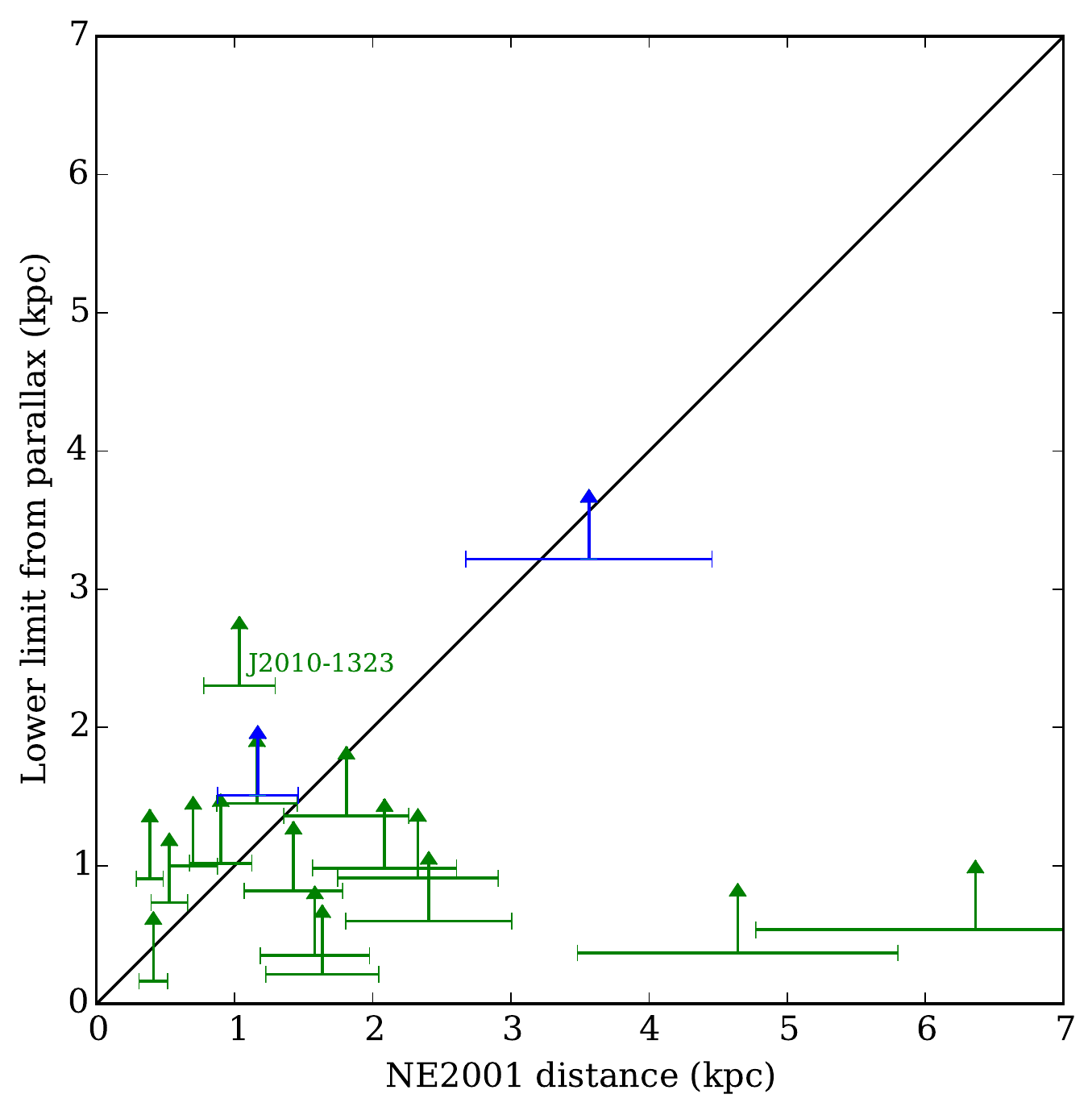}
  \caption{Lower limits on distances from parallax measurements versus distances from the NE2001 dispersion measure model for pulsars without significant parallax detections in our data set (lower part of Table~\ref{tab:pxlimits}).  Lower limit parallax distances are plotted as 95\% confidence values; NE2001 dispersion model distances are plotted with 25\% uncertainties. Previous parallax measurements or limits for PSRs~B1855+09 and B1937+21 were used as input data to the NE2001 model; our limits for these pulsars are shown in blue.  All of our other limits are shown in green.
  \label{fig:pxlimitvsdm}
}
\end{figure}

In Figures \ref{fig:sigpxvsdm} and \ref{fig:pxlimitvsdm}, we compare our measured distances and limits with the predictions of the NE2001 Galactic electron density model \citep{CordesLazio2002} given the measured DMs of these pulsars.  Note that previous parallax measurements or limits for PSRs~J1713+0747, J1744-1134, B1855+09, and B1937+21 were used as input data to the NE2001 model, generally with much larger uncertainties than in the present work (J. Cordes, private communication).

We find that, in large part, distances predicted based upon dispersion measure and the NE2001 model were in agreement with parallax derived distances and limits.  Among the pulsars for which we have distance measurements (not just limits), there were only two pulsars for which there is no possible distance that is both within the $2\sigma$ parallax distance measurement range and within $25\%$ of the NE2001 distance prediction, PSRs J1614$-$2230 and J1909$-$3744 (Figure~\ref{fig:sigpxvsdm}).  In the case of parallax distance lower limits, the NE2001 electron density model fared equally as well. Apart from a few outlying pulsars, the limits were not too far from the distance derived via dispersion measure and the NE2001 model.

We calculated the dispersion measures for which the NE2001 model yields the distance measurements we derived from parallax. For PSR J1614$-$2230, the dispersion measure required for agreement would shift from $34.5$ to $10.8~{\rm pc\,cm}^{-3}$, and for PSR J1909$-$3744 it would shift from $10.4$ to $34.4~{\rm pc\,cm}^{-3}$.

\begin{figure}
\centering
\includegraphics[scale=0.5]{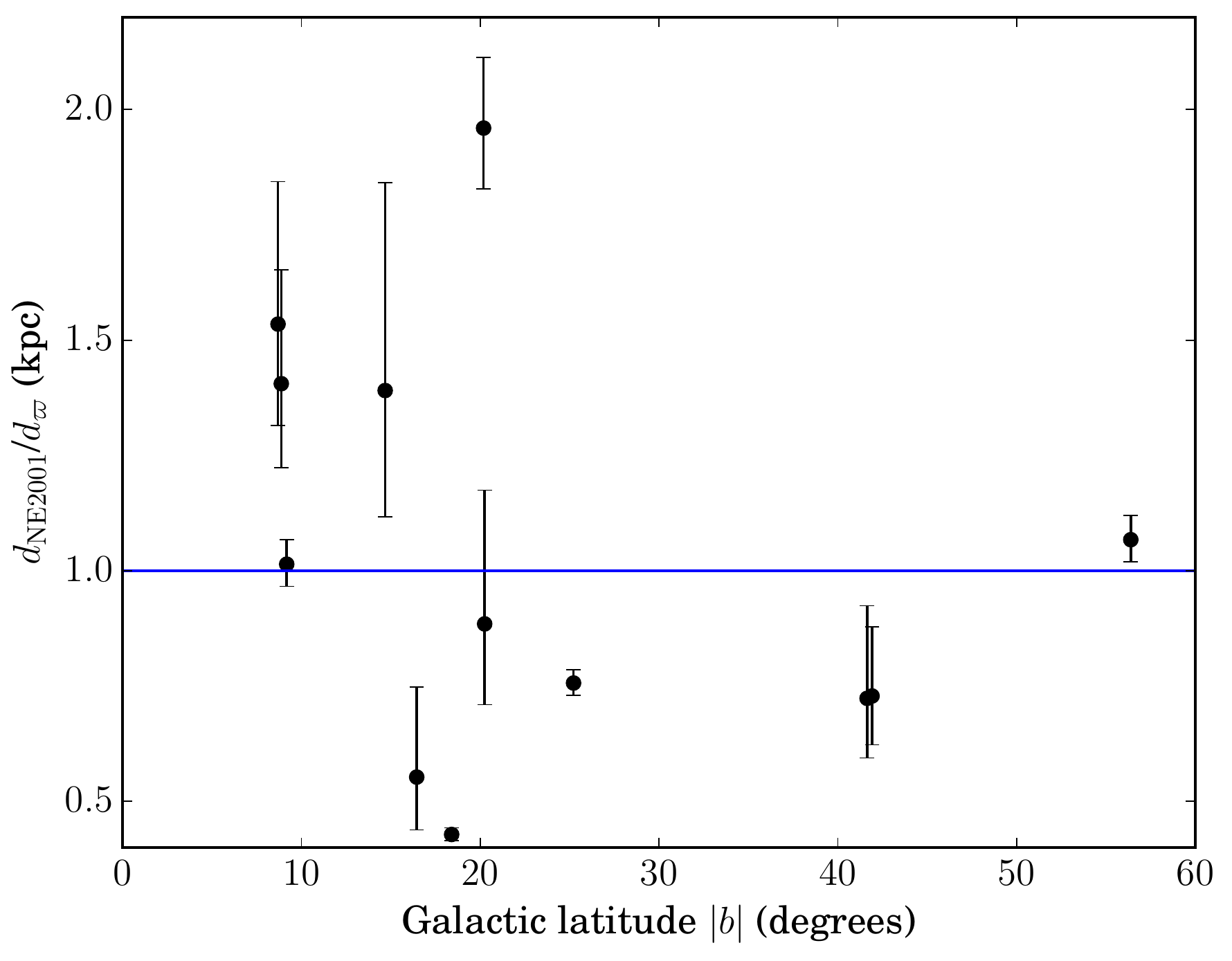}
\caption{Comparison of distances derived from dispersion measure via the NE2001 electron density model and parallax distances as a function of absolute Galactic latitude.
\label{fig:pxdm_gb}
}
\end{figure}

Recent work on the Galactic electron distribution can be found in, e.g., \cite{Cordes2013}, \cite{Schnitzeler2012}, and references therein.  New and refined pulsar distance measurements can contribute to this effort.  For example,
\cite{Chatterjee2009} reported that NE2001 underestimated distances to some pulsars at high Galactic latitudes, implying the need for a larger disk scale height or disk electron density \citep[see also][]{Schnitzeler2012}. In Figure \ref{fig:pxdm_gb}, we plot $d_{\rm NE2001}/d_{\varpi}$ vs. Galactic latitude. No trends are apparent in our measurements, though we have fewer high Galactic latitude sources than \cite{Chatterjee2009} and are 
limited by small number statistics.
For those pulsars with dispersion distances that did not agree within two standard deviations of the parallax-derived measurements, we find no correlation between their sky positions. The sources that disagree are at varying Galactic longitudes, although all are greater than $18^{\circ}$ from the Galactic plane.

\section{Millisecond pulsar population kinematics}
\label{sec:mspvelocities}
\makeatletter{}\subsection{Galactocentric velocity components}

To analyze the kinematics of the millisecond pulsars, we used the measured proper motions and distances to calculate velocity vectors in galactocentric coordinates at the pulsars' standards of rest.  Because there are not line-of-sight velocity measurements for most pulsars, complete three-dimensional velocity vectors are not obtainable, and our analysis for such pulsars uses only velocity components that are close to orthogonal to the line-of-sight; we describe this further below.

For this analysis, we used both measurements from the present work and previously reported measurements found in the literature.  We included all millisecond pulsars for which proper motions were available with 5$\sigma$ significance in two orthogonal components.  When available, we used our proper motion measurements (Table~\ref{tab:propmo}); for PSRs~J1012+5307, J1738+0333, and J1949+3106, although they are included among our measurements, higher-precision measurements are available in the literature.  For these and other pulsars, we used the values listed in Table~\ref{tab:propmo_nonNANOGrav}.

For distances, we used measured parallax distances with 3$\sigma$ or greater significance from our measurements (Table~\ref{tab:pxlimits}) or from the literature (Table~\ref{tab:px_nonNANOGrav}).  For pulsars without parallax distances, we estimated distances from dispersion measures and the NE2001 electron density model.

For each pulsar, we calculated a two-dimensional velocity in the reference frame of the Sun and transformed it to the galactocentric coordinates at the location of the pulsar.  For this transformation, we used solar motion $U_{\odot} = 11.1 \mbox{ \kms} $, $V_{\odot}=12.24 \mbox{ \kms} $, and $W_{\odot} = 7.25 \mbox{ \kms} $, \citep{Schonrich2010}, and we used
solar galactocentric distance $R_{\odot} = 8.34 \mbox{ kpc}$ and Galactic rotation velocity $\Theta_0=240 \mbox{ \kms}$ \citep{Reid2014}. 

The line-of-sight (LOS) velocity has been measured via optical observations for three pulsars of interest (Table~\ref{tab:los_velocity}), allowing three-dimensional galactocentric velocities to be calculated for these pulsars.  For the remaining pulsars, we calculated and removed the line-of-sight velocity expected for the pulsar if at rest in its standard of rest.  This LOS correction is only a reasonable approximation for components of the galactocentric velocity that are nearly perpendicular to the LOS. In the calculations below, we only include measurements of galactocentric velocity components that are at least $70^\circ$ from the direction of the LOS to the pulsar.

\subsection{Dispersion of millisecond pulsar velocities}

\begin{figure*}
 \centering
 \includegraphics[scale=0.7]{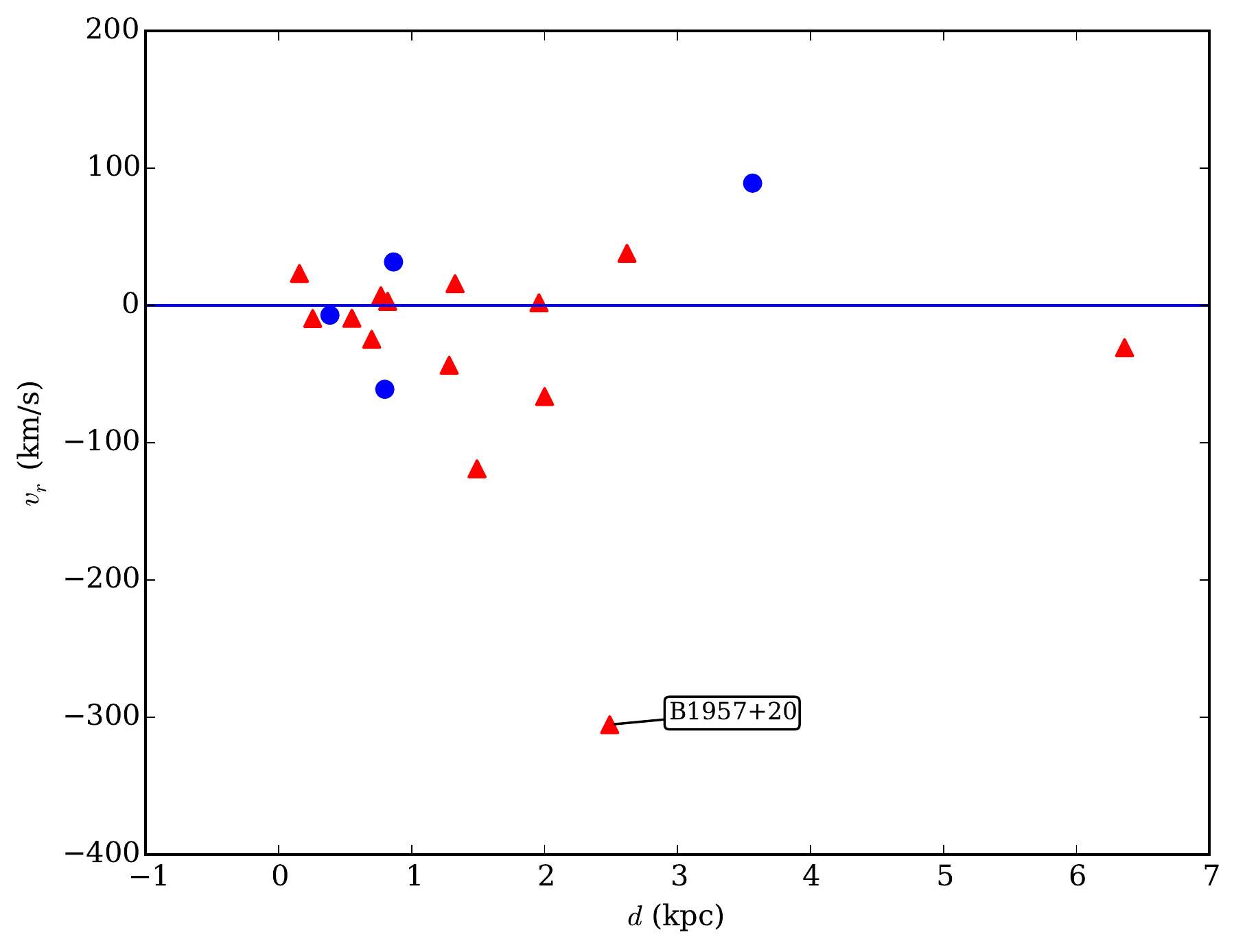}
 \caption{Millisecond pulsar radial-velocities versus distance for pulsars whose radial velocity component was within $20^{\circ}$ of perpendicular to the line-of-sight. Isolated millisecond pulsars are shown as blue circles, while millisecond pulsars in binary systems are shown as red triangles.}
 \label{fig:rv_los}
\end{figure*}

\begin{figure*}
 \centering
 \includegraphics[scale=0.7]{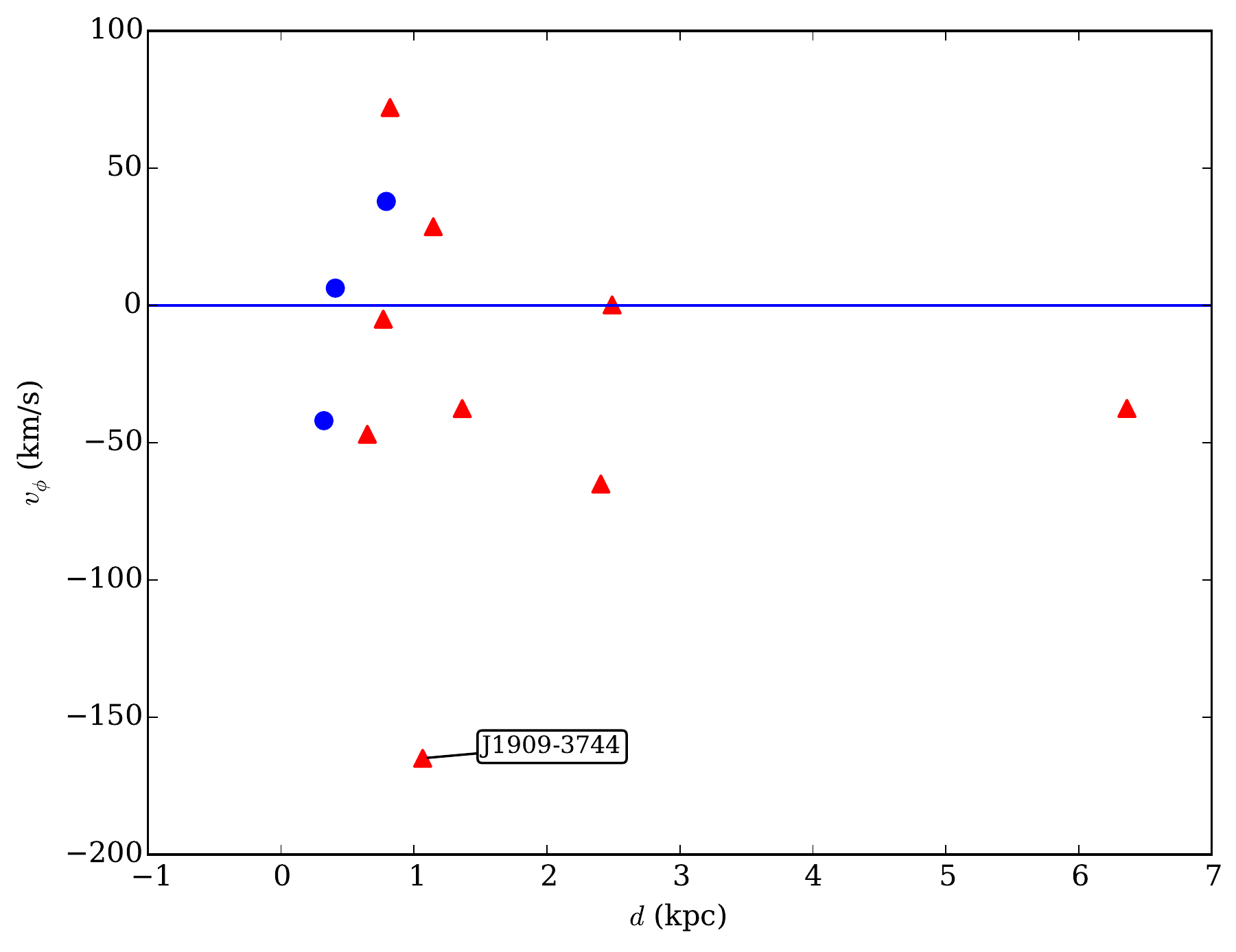}
 \caption{Millisecond pulsar azimuthal-velocities versus distance for pulsars whose azimuthal velocity vector is within $\
20^{\circ}$ of perpendicular to the line-of-sight. Isolated millisecond pulsars are sh\
own as blue circles, while millisecond pulsars in binary systems are shown as r\
ed triangles.}
 \label{fig:azimuthv_los}
\end{figure*}

\begin{figure*}
 \centering
 \includegraphics[scale=0.7]{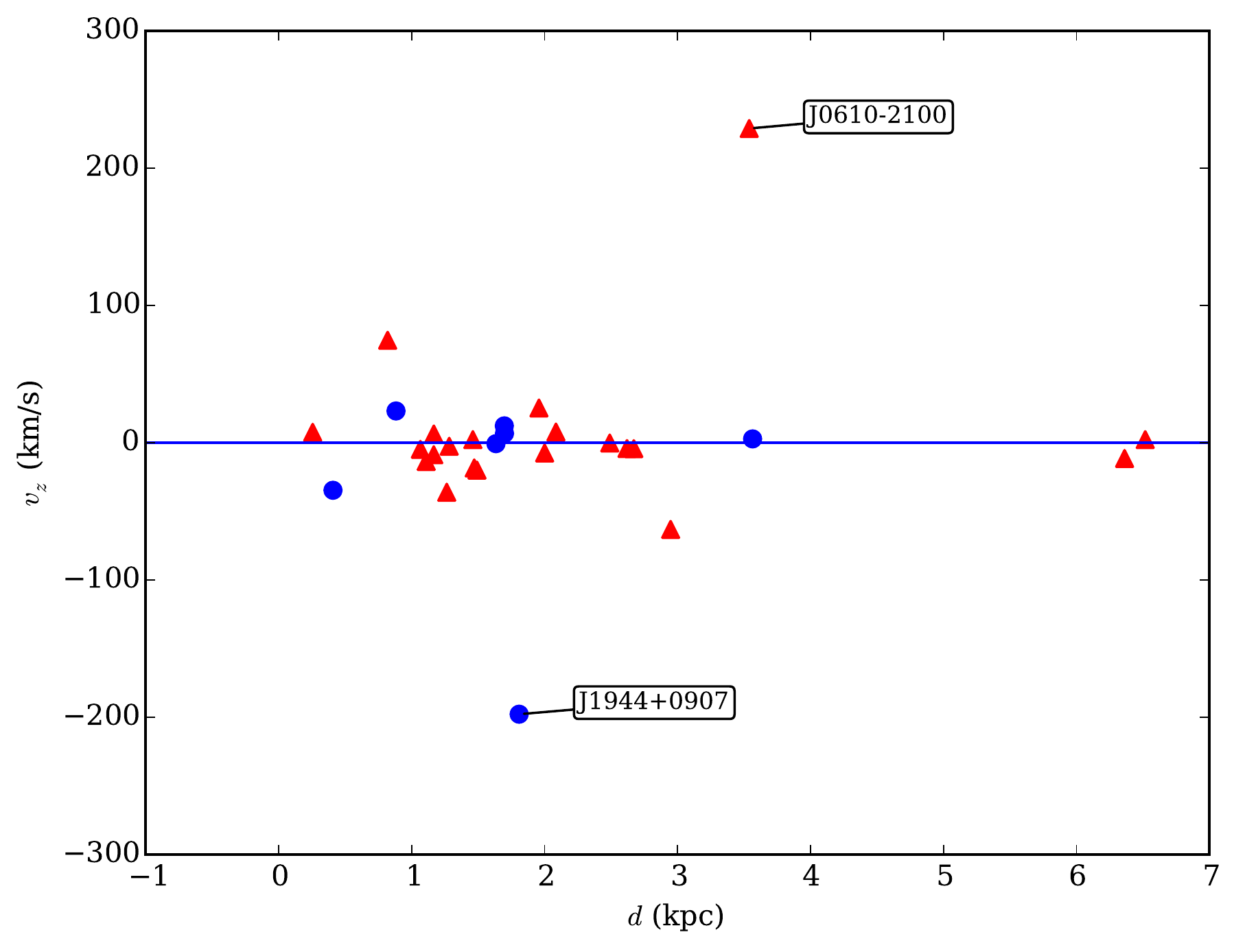}
 \caption{Millisecond pulsar z-velocities versus distance for pulsars  within $20^{\circ}$ latitude to the Galactic plane. Isolated millisecond pulsars are shown as blue circles, while millisecond pulsars in binary systems are shown as red triangles.}
 \label{fig:zv_los}
\end{figure*}

Using the algorithm and criteria describe above, we calculated Galactic radial velocities of
18 pulsars (Figure~\ref{fig:rv_los}), azimuthal velocities of 12 pulsars (Figure~\ref{fig:azimuthv_los}), and perpendicular velocities of 28 pulsars (Figure \ref{fig:zv_los}).  The component values included in each velocity dispersion calculation are listed in Table~\ref{tab:velocity}.

Using the values in Table~\ref{tab:velocity}, 
we find the dispersions of these velocity measurements to be:
\begin{gather}
\label{eqn:disp_all}
\sigma_R = 81 \mbox{ \kms}
\hspace*{12pt}
\sigma_{\phi} = 58 \mbox{ \kms}
\hspace*{12pt}
\sigma_z = 62 \mbox{ \kms}. 
\\
\mbox{(All measurements)} \notag
\end{gather}
Visual examination suggests that there are four outliers in the velocity component distributions: PSRs B1957+20 ($v_R$, Figure~\ref{fig:rv_los}); J1909$-$3744 ($v_{\phi}$, Figure~\ref{fig:azimuthv_los}); and J1944+0907 and J0610-2100 ($v_z$, Figure~\ref{fig:zv_los}).  We used the median absolute deviation as a robust method, under the assumption of a Gaussian velocity distribution, to test whether these points are outliers \citep{Leys2013,NumericalRecipes}.   For each pulsar, we calculated $|v-M|/\mbox{MAD}$, where $v$ is the velocity component of the pulsar of interest, $M$ is the median of all such measurements, and MAD is the median absolute deviation.  We obtained values of $7.2$, $3.6$, $14.2$, and $16.7$ for PSRs~B1957+20, J1909$-$3744, J1944+0907, and J0610-2100, respectively. Based on the criterion $|v-M|/\mbox{MAD}>3$, we argue that these are indeed outliers. 
We discuss possible physical mechanisms for such outliers in Section \ref{sec:conclusion}. 
If we exclude these three outlier pulsars from all velocity dispersion calculations, the dispersions of the velocity measurements are:
\begin{gather}
\label{eqn:disp_exclude}
\sigma_R = 46 \mbox{ \kms}
\hspace*{12pt}
\sigma_{\phi} = 40\mbox{ \kms}
\hspace*{12pt}
\sigma_z = 24 \mbox{ \kms}. 
\\
\mbox{(Excluding outliers)} \notag
\end{gather}
Using these dispersions, PSR~B1957+20 is $6.1$ sigma from the mean, PSR~J1909$-$3744 is $3.6$ sigma from the mean, PSR~J1944+0907 is $8.2$ sigma from the mean, and PSR~J0610-2100 is $9.5$ sigma from the mean, reinforcing their status as outliers.

A possible concern with the velocity dispersion measurements is selection effects due to non-uniform sky coverage of pulsar surveys.  Search programs such as the PALFA \citep{PALFA} and PMB \citep{Lorimer2006} surveys have concentrated on low Galactic latitudes.  As an old population, millisecond pulsars found close to $b = 0^{\circ}$ are more likely to have smaller velocities in the direction away from the plane. Thus, surveys focusing on the Galactic plane might be susceptible to biased velocity distributions.  However, there is no similar reason to expect azimuthal or radial velocity distributions to be susceptible to such selection effects, so the latter components are more robust measurements of the true underlying kinematics of the millisecond pulsar population.

Velocity dispersions of other stellar populations in the Galaxy are observed to have $\sigma_R>\sigma_{\phi}>\sigma_z$, with typical ratios $\sigma_z/\sigma_R\simeq 0.5$ and $\sigma_\phi/\sigma_R\simeq 0.55\ {\rm to}\ 0.7$ \citep{GalacticAstronomy}.  The velocity dispersions of equation~\ref{eqn:disp_exclude} (outliers excluded) follow this pattern, suggesting that these calculated dispersions are relatively free of bias due to pulsar search selection effects and hence represent the intrinsic velocity dispersions of this population.  The velocity dispersions of equation~\ref{eqn:disp_all} (all measurements) don't strictly follow these patterns: they have $\sigma_R>\sigma_{\phi}\simeq\sigma_z$, and $\sigma_z/\sigma_R$ is a bit higher than for typical stellar populations.  However, this could easily be a case of small-number statistics: removing a single high-velocity source from the $\sigma_z$ calculation would bring these numbers into alignment with other stellar populations.

\subsection{Isolated vs. binary millisecond pulsars}

Previous works have studied the velocities distributions for isolated and binary millisecond pulsars and found no significant difference \citep{Gonzalez2011}. To test whether there exists a difference in velocities between isolated and binary millisecond pulsars with our new proper motion values, we calculated the radial, azimuthal and perpendicular velocity dispersions for each subpopulation, recognizing that small number statistics allow only a rough comparison.  We obtained velocity dispersions of 55 and 43~\kms\ for the radial velocity dispersions of isolated and binary millisecond pulsars, respectively.  We obtained dispersions of 33 and 42~\kms\ for the isolated and binary azimuthal velocity dispersions and 18 and 25~\kms\ for the isolated and binary perpendicular velocity dispersions.  In all cases the similarities between the isolated and binary dispersion values imply there is little to no difference between the two subpopulations.

These calculations excluded the outliers described above.  
Including the outliers for each velocity component significantly changes the dispersions.  Since the binary and isolated dispersions are calculated separately, including an outlier only affects either isolated or binary pulsar dispersion for a given component.  For the radial component, the binary dispersion becomes 87~\kms. For the azimuthal component, the binary dispersion becomes 63~\kms. For the perpendicular component, the isolated dispersion becomes 72~\kms and the binary dispersion becomes 55~\kms.  The consistency between the binary and isolated dispersions when the outliers are excluded, and the inconsistency (with no underlying pattern) when the outliers are included lends itself well to arguing that these pulsars are indeed outliers.
 
\subsection{Comparison of millisecond pulsar velocity dispersion with other populations}

Millisecond pulsars are an old population.  Characteristic spin-down ages,
$\tau=P/2\pdot$, of the pulsars in the NANOGrav nine-year data set calculated
using the observed spin parameters range from 0.2 to 29~Gyr, with a median
of 5~Gyr.  Correcting for kinematic effects to estimate intrinsic spin-down
ages (\S\ref{sec:pdotdistances}) gives larger ages: 0.2 to 46~Gyr with a
median of 8~Gyr for the 30 of our pulsars with more than two years of data
(and hence well-measured proper motions, as needed for the kinematic
correction).  Obviously, pulsars cannot be older than a Hubble time; some,
perhaps all of these ages must be overestimates, which is easily explained if
millisecond pulsar spin periods at formation are not much shorter than their
present-day spin periods \citep[e.g.,][]{KiziltanThorsett2010,TaurisLangerKramer2012}.  
Nevertheless, it seems reasonable to assume that 
millisecond pulsars are typically at least several Gyr old.  We will use
ages of $\tau\sim 5$ and 10~Gyr in the comparisons below.

Velocity dispersions of stellar populations are well known to correlate with
age \citep[e.g.,][]{GalacticAstronomy,Dehnen1998}, with younger stars described as a thin
disk and older stars described as a thick disk, although the mechanism behind
the correlation is not well established \citep[e.g.,][]{Sharma2014}.  \cite{Cordes1997}
pointed out that the diffusion of millisecond pulsars into a thick disk contributes
a significant portion of their observed velocities.  Here we
consider our millisecond pulsar velocity dispersion measurements in the context of dispersion-age
relations developed from studies of optical stellar populations.  
A caveat to this comparison is that optical stellar population
modeling tends to use stars in the solar neighborhood, whereas our millisecond
pulsar population is spread over a larger volume.   Nevertheless, we have
already seen that the ratios $\sigma_R:\sigma_\phi:\sigma_z$ for millisecond
pulsars follow those of other stellar populations, so it seems plausible that the
magnitudes of the dispersions may be comparable as well.

\cite{Aumer2009} use local stellar data to fit equations of the form
  $\sigma(\tau) = v_{10}[(\tau+\tau_1)/(10~{\rm Gyr}+\tau_1)]^{\beta}$
for each component of velocity dispersion.  For radial, azimuthal, and
perpendicular components, they found $\beta=0.307$, 0.430, and 0.445;
$\tau_1=0.001$, 0.715, and 0.001; and $v_{10}=41.899$, 28.823, and 23.831~\kms,
respectively.  For an age of $\tau\sim 5~{\rm Gyr}$, this gives
$\sigma_R=34$~\kms, $\sigma_\phi=22$~\kms, and $\sigma_z=18$~\kms.  For an age
of $\tau\sim 10~{\rm Gyr}$, it gives $\sigma_R=42$~\kms,
$\sigma_\phi=28$~\kms, and $\sigma_z=24$~\kms.

\cite{Dawson2010}, fit an equation to binned $z$-velocity vs.\ age data and find
an empirical relation $\sigma_z = 10.1(1+\tau/{\rm Gyr})^{0.45}$~\kms.  For
ages of $\tau\sim 5$ and 10~Gyr, this gives $\sigma_z\sim 23$~\kms and
30~\kms respectively.

For typical millisecond pulsar ages, then, the velocity dispersions from the model
of \cite{Aumer2009} are only modestly different than our velocity dispersions in the
measurements that exclude outliers (Equation~\ref{eqn:disp_exclude}), and the
velocity dispersions of \cite{Dawson2010} are a nearly perfect match.

We conclude that, if this characterization of millisecond pulsar kinematics as
a Gaussian velocity distribution with a small number of outliers is correct,
then the bulk of these objects need essentially no velocity boost to reach
their observed velocities, since they are comparable to other stars of similar
ages.  It is generally accepted that most neutron stars receive a kick at
birth \citep[e.g.,][]{Hobbs2005}.  When the neutron star has a stellar-mass
companion, the resulting space motion of the center of mass will generally be
smaller than in the case of an isolated neutron star due to the mass of the
binary.  However, the observed outliers in our sample could be the result of
fortuitously directed kicks that produce significant center-of-mass
velocities.  Furthermore, it has been proposed that O-Ne-Mg-core stars undergo
electron capture supernovae with small velocity kicks, whereas iron-core stars
undergo supernovae with large velocity kicks
\citep{Podsiadlowski2004,VanDenHeuvel2011,Tauris2013} .  This dichotomy may
also contribute to our observed velocity distribution.

On the other hand, if the outlier model is incorrect, and the velocity
dispersions in Equation~\ref{eqn:disp_all} are a more appropriate measure of
the millisecond pulsar population, then their velocities are moderately larger
than other stars of similar ages, but a significant portion of their
velocities must still be attributable to the same mechanism that increases
other stellar velocities over time (whatever that mechanism is).

In either model, unlike canonical pulsars, millisecond pulsar velocities are very
low, and require small velocity boosts at most during their formation.

\section{Conclusion}
\label{sec:conclusion}
\makeatletter{}
We have measured and refined distances to twelve millisecond pulsars and found distance limits on eighteen more (Table \ref{tab:pxlimits}). These distances will find uses in a variety of applications, from the physics of pulsars themselves (e.g., the distance is needed if using a measured flux density to calculate its luminosity), to the analysis of the ionized interstellar medium. In \S\ref{sec:ne2001} we focused on the latter application and found the distances predicted by the NE2001 electron density model in general agreement with those calculated from parallax. 

For 7 of the 37 NANOGrav millisecond pulsars, we also calculated distance or a $95\%$ upper limit on distance from the change in orbital period, $\pbdot$. This independent calculation of distance is in good agreement with the distances derived from parallax. As an additional independent method of calculating distances, we derived upper limits on distance for 17 of the NANOGrav millisecond pulsars from the rotational spin down rate, $\pdot$. Apart from PSR J1024$-$0719 (\S\ref{app:J1024}), these upper limits are in agreement with the parallax distances and $95\%$ lower limits. 

We have measured proper motions of 35 millisecond pulsars.  We used 30 of
these (from pulsars observed for more than two years), in combination with
distance estimates and measurements found in the literature, to analyze pulsar
motion in Galactocentric coordinates.  We found the velocity dispersion
components to follow similar proportions as other stellar populations
(Equations~\ref{eqn:disp_all} and~\ref{eqn:disp_exclude}).  We propose two
mathematical models of the velocity dispersion magnitudes.  In one model, the
bulk of the millisecond pulsar population has essentially the same velocity
distribution as other stellar populations with ages of several Gyr, and
a small number of pulsars are high-speed outliers.  We speculate that the
outlier velocities could be due to fortuitously directed kicks during
neutron-star formation, though we cannot rule out either a different
formation mechanism than the bulk of the millisecond pulsar population
or an origin in a different dynamical population (e.g., halo stars).
  In the other model, the millisecond pulsars come from a single velocity 
distribution, which has dispersion much smaller than canonical pulsars, but moderately larger than
other stellar populations with ages of several Gyr.  In this model, the
pulsar velocities derive from a combination of their dynamical origin as thick
disk objects and from modest velocity boosts during millisecond pulsar formation, 
presumably at the time they were formed as neutron stars in supernovae.

Our goal has been to develop an empirical description of millisecond pulsar
dynamics based on measured pulsar parameters alone.  A more comprehensive
study would take into account the directions and sensitivities of pulsar
search programs, the luminosity distribution of millisecond pulsars, 
millisecond pulsar birth locations and dynamical evolution, Lutz-Kelker bias, 
uncertainties in the dispersion distance model, and so on.  Such studies
are typically done via Monte Carlo simulations; see \cite{Lorimer2013}
for an overview and further references.  A conclusion from the present work
is that the analysis of velocities in such studies should take into account
the significant dispersion in millisecond pulsar velocities due simply to
their large ages, and that the separation of velocities into galactocentric
components (not just total velocities) is important.

\acknowledgements 

{\it Author Contributions.}  
The paper evolved from the
undergraduate senior thesis of AMM, supervised by DJN.  AMM undertook the bulk
of the calculations; generated most of the text, figures, and tables; and
coordinated the development of the paper.  DJN
contributed additional text, figures, tables, and analysis, including
the appendix on PSR~J1024$-$0719.  EF contributed text and
calculations to the binary distance analysis section.  
This paper is based on positions, proper motions, and parallaxes derived from the
NANOGrav nine-year data set \citep{Arzoumanian2015}.  ZA, KC, PBD, TD, RDF, EF,
MEG, GJ, MJ, MTL, LL, MAM, DJN, TTP, SMR, IHS, KS, JKS, and WWZ all ran
observations and analyzed timing models for this data set.  
Additional specific contributions to the data set are summarized in
\cite{Arzoumanian2015}; here we note the particularly critical contributions
of PBD, who generated the TOAs and developed novel techniques for removing
systematics from the data, and JAE, who developed and implemented the noise
model used in the timing analysis.

We thank S.\ Chatterjee, J.\ Cordes, J.\ Verbiest, J.\ Lazio, and C.\ Ng 
for useful discussions and comments on the manuscript.
The ATNF pulsar catalogue\footnote{\url{http://www.atnf.csiro.au/people/pulsar/psrcat/}}
\citep{ATNF}, was an invaluable resource for this project, as was S.\ Chatterjee's
online pulsar parallax list\footnote{\url{http://www.astro.cornell.edu/~shami/psrvlb/parallax.html}}.
The NANOGrav project receives support from National
Science Foundation (NSF) PIRE program award number 0968296 and NSF
Physics Frontier Center award number 1430284.  NANOGrav research at UBC
is supported by an NSERC Discovery Grant and Discovery Accelerator
Supplement and the Canadian Institute for Advanced Research.  Part of
this research was carried out at the Jet Propulsion Laboratory,
California Institute of Technology, under a contract with the National
Aeronautics and Space Administration.  TP was a student at
the National Radio Astronomy Observatory while this project was
undertaken.  MTL was partially supported by NASA New York Space Grant 
award number NNX15AK07H. The National Radio Astronomy
Observatory is a facility of the NSF operated under cooperative
agreement by Associated Universities, Inc.  The Arecibo Observatory is
operated by SRI International under a cooperative agreement with the NSF
(AST-1100968), and in alliance with Ana G.  M\'{e}ndez-Universidad
Metropolitana, and the Universities Space Research Association.

\bibliographystyle{apj}

\appendix

\makeatletter{}\section{PSR J1024$-$0719}
\label{app:J1024}

As described in~\S\ref{sec:pdotdistances}, we have derived two distance limits for
PSR~J1024$-$0719 that contradict one other: an upper limit on distance from
spin-down, $d_{\dot{P}}<0.427$~kpc, and a lower limit on distance from
parallax, $d_{\varpi}>0.91$~kpc.   To quantify this discrepancy, note 
that if the pulsar lies within the bound established by the spin-down distance, 
then the parallax measurement, $\varpi=0.6\pm 0.3$~mas is in error by at least
$6\sigma$.  The resolution of this conflict is not clear.  In this appendix,
we summarize the reasons the spin-down distance measurement is
robust, we describe tests of our parallax measurement, we summarize other
observations of this pulsar, and we present an orbital model which
is a candidate for resolving this discrepancy.

PSR~J1024$-$0719 is isolated (non-binary) and has rotation properties typical
of millisecond pulsars.  It has the largest proper motion among the sources in
this paper, and one of the smallest dispersion measures.

\subsection{Spin-down distance}

In \S\ref{sec:pdotdistances}, the spin-down distance was calculated from 
equation~\ref{eqn:pdot2} assuming
the observed spin-down, $\pdot=(1.8551\pm0.0001)\times 10^{-20}$, is entirely due to
kinematic effects.  Similar spin-down distance upper limits have been
previously calculated for this pulsar \citep{Toscano1999,Espinoza2013}.

At the upper limit distance, 0.427~kpc, the contributions to $\pdot$
of the three terms of the right hand side of equation~\ref{eqn:pdot2} are
$-0.036\times 10^{-20}$ for the $a_z$ term; $-0.009\times 10^{-20}$ for the
Galactic rotation term; and $1.900\times 10^{-20}$ for the proper motion term,
where we have used the pulsar period $P=5.162$~ms.
The proper motion term dominates the calculation, so the calculation is
robust against uncertainties in the Galactic potential and rotation and we can
write, with error of no more than 2\%, $d_{\pdot}\lesssim \pdot c/P\mu^2$.
As shown in Table~\ref{tab:1024_previous}, our measurements of $\pdot$ and $\mu$
are in agreement with two previously published timing solutions for this
pulsar.  For these reasons, this upper limit determination is robust, as
long as the pulsar is spinning down.

\subsection{Parallax measurement}

The distance upper limit from $\pdot$\ implies a parallax lower limit of 
$\varpi>d_{\pdot}^{-1}=2.3$~mas.

Our parallax measurement for PSR~J1024$-$0719 is $\varpi=0.6\pm 0.3$~mas.  There
was one previously reported measurement, $\varpi=1.9\pm 0.9$~mas \cite{Hotan2006}, 
which differs by $1.5\sigma$ from our value.  While this
previous value is compatible with the parallax implied by the $\pdot$ measurement,
it has a large uncertainty and is only marginally significant.

To test the validity of our parallax measurement, we performed a series of
tests on the data set as detailed below and as summarized in
Table~\ref{tab:1024_px}.  For each test, the table shows the best-fit
parallax; the $\chi^2$ of the fit; the number of degrees of freedom, $n_{\rm
dof}$; and the reduced $\chi^2$.  Each of our tests used the noise model
values for this source as determined by \cite{Arzoumanian2015}, which consists
only of white-noise terms.
In all tests, we found the parallax to be
consistent with our standard measured value and to be lower than that implied
by the $\pdot$\ distance limit.

{\it Data subset tests.}
Our data set consists of measurements taken with two instruments,
GASP and GUPPI (\S\ref{sec:obs}).  GASP data were collected at three epochs, 
in years 2009.8 through 2010.1, and GUPPI data were collected
at 49 epochs, 2010.2 through 2013.8.  (There are also a very small number
of GASP TOAs within the GUPPI date range covering frequencies without good
GUPPI TOAs.)
We ran independent timing solutions on (i) the GUPPI data only; (ii) the first half
of the data set, 2009.8 through 2011.8; and (iii) the second half of the data set,
2011.8 through 2013.8.

{\it Timing noise tests.}
\cite{Arzoumanian2015} found no evidence for red noise in this data set.  
Nevertheless, as a test, we ran independent timing solutions with extra
spin frequency derivatives as a proxy for timing noise.  Defining
$f_i\equiv d^if/dt^i$, we ran tests fitting for all the usual parameters
and additionally (i) $f_2$; (ii) $f_2$ and $f_3$; (iii) $f_2$ through $f_6$.

{\it Dispersion measure tests.}  Our standard solution fits for independent
values of dispersion measure at every epoch.  These dispersion measure values
are dominated by a linear trend \citep[Fig.~14]{Arzoumanian2015}.  We ran
two tests in which we fit dispersion measure as a linear trend combined with
a solar wind electron density with a $1/r^2$ falloff, with electron density
at 1~AU of $n_{0,{\rm solar}}$ of 0 and 10~cm$^{-3}$ in the two tests.  In such
models, the solar wind electron density is highly covariant with the best-fit
parallax value, so in principle one can adjust $n_{0,{\rm solar}}$ to attain
any derived parallax value; however, we found that the quality of the fit 
diminished significantly if $n_{0,{\rm solar}}$ was increased beyond 10~cm$^{-3}$.

\subsection{Other observations of this pulsar}
\label{sec:1024_other}

\cite{Espinoza2013} noted that the gamma-ray emission of this pulsar would be
unusually high at this distance (they used 0.410\,kpc), and that a closer
distance (0.350~kpc) would be needed for its gamma-ray luminosity to be
similar to other millisecond pulsars.  This pulsar has also been detected
in X-rays \citep{Espinoza2013,Zavlin2006}.

\cite{Sutaria2003} presented optical observations of the field of 
PSR~J1024$-$0719.  They detected two sources near the pulsar.
One was bright ($U=22.11$, $V=19.82$, $R=18.89$, $I=18.17$), with a spectrum
similar to a K-type dwarf star.  One was faint ($U=23.8$, $V=24.9$, $R=24.4$, $I=24.2$).
The bight star may have a proper motion in a direction similar to the pulsar,
although uncertainties are large.  Since the pulsar is isolated, the
bright source may be unassociated with the pulsar, but its presence
is an interesting coincidence.

The NE2001 electron density model predicts a distance of $d_{\rm DM}=0.39$~kpc
(Table~\ref{tab:pxlimits}).

\subsection{Discussion}

Here we speculate what circumstances could reconcile the measurements
if the pulsar is at the parallax lower limit distance, $d_\varpi=0.91$~kpc.
At this distance, according to equation~\ref{eqn:pdot2}, the observed
period derivative would be biased upward by $4.0\times 10^{-20}$, 
i.e., this is a lower limit to the observed
period derivative if equation~\ref{eqn:pbdot2} fully describes the biases to
the observed period derivative.  Since the actual observed
period derivative is $1.8\times 10^{-20}$, under this model, 
the terms of equation~\ref{eqn:pbdot2} are not sufficient to explain
the observed value.
An additional bias of $\pdot_{\rm bias}=2.2\times 10^{-20}$ is needed,
which could arise from an additional acceleration 
of order $a/c=\pdot_{\rm bias}/P=4\times 10^{-18}~{\rm s}^{-1}$.  

Such an acceleration could be caused by the potential of a globular
cluster---indeed, millisecond pulsars in globular cluster cores have
a wide range of observed positive and negative $\pdot$ values.  However,
there is no cluster in the direction of PSR~J1024-$0719$.

Such an acceleration could also be caused by binary motion in a wide orbit.
The acceleration would change over the course of the orbit, causing
a change in the observed $\pdot$, i.e., a nonzero $\ddot{P}=d\pdot/dt$.  
The difference in observed $\pdot$ between \cite{Verbiest2009} and 
the NANOGrav nine-year value is $\Delta\pdot=(3\pm8)\times 10^{-23}$ over a time
span of 9.6~yr (using the centers of the observing data spans); this
corresponds to an approximate upper limit $\ddot{P}\lesssim 1\times 10^{-23}~{\rm yr}^{-1}$.
An orbit would have to involve $\pdot$ values that varied on the scale of
$\pdot_{\rm bias}=2.2\times 10^{-20}$, so the time scale of such variations
would be $\Delta t \gtrsim \pdot_{\rm bias}/\ddot{P}=2000$~yr, or an orbital
period of $T\simeq 2\pi\Delta t \gtrsim 14000$~yr.  For acceleration $a=\omega^2 r$,
where $\omega=2\pi/T$, this gives pulsar orbital radius 
$r\gtrsim 2\times 10^{4}~{\rm s}=40~{\rm AU}$.  For a 1.4~M$_\odot$ pulsar, this
would require a companion star of mass $m_2\gtrsim 0.1~{\rm M}_\odot$.  Much larger
masses would also satisfy the constraints; intriguingly, this includes the mass of a K-type
star as observed by \cite{Sutaria2003} (\S\ref{sec:1024_other}).

Placing the pulsar at this large distance would present challenges, though.
Its two-dimensional space velocity derived from proper motion 
would be 260~\kms, higher than typical
millisecond pulsars; its DM would be much lower than that predicted
by the NE2001 electron density model; and its gamma-ray efficiency would 
be very high, much greater than the value calculated in~\cite{Espinoza2013}.

\clearpage

\makeatletter{}\begin{deluxetable*}{lcccllc}
\tablewidth{0pc}
\tabletypesize{\small}
\tablecaption{Positions\tablenotemark{a}\label{tab:equatorial}}
\tablehead{\colhead{PSR} & \multicolumn{2}{c}{Ecliptic Coordinates}                      && \multicolumn{2}{c}{Equatorial Coordinates}          & \colhead{Epoch (MJD)\tablenotemark{b}} \\
\cline{2-3}\cline{5-6}
\rule{0pt}{10pt} & \colhead{$\lambda$ ($^\circ$)}    & \colhead{$\beta$ ($^\circ$)}  && \colhead{$\alpha$ (hh:mm:ss)} & \colhead{$\delta$ (dd:mm:ss)} &  
}
\startdata
J$0023+0923 $ & $ \phn\phn     9.07039784(4)         \phn $ & \phs \phn $      6.3091086(3)         \phn\phn $  &&	 	00:23:16.87910(3)  &  	\phn09:23:23.871(1)   & $ 56179 $ \\
J$0030+0451 $ & $ \phn\phn     8.91035630(1)         \phn $ & \phs \phn $      1.4456962(5)         \phn\phn $  &&	 	00:30:27.42826(5)  &  	\phn04:51:39.711(2)   & $ 54997 $ \\
J$0340+4130 $ & $     \phn    62.61406221(5)         \phn $ & \phs      $     21.3344746(2)         \phn\phn $  &&	 	03:40:23.28818(2)  &  	\phn41:30:45.2903(5)  & $ 56279 $ \\
J$0613-0200 $ & $     \phn    93.79900655(2)         \phn $ &           $   -25.40713269(4)             \phn $  &&	 	06:13:43.975631(3) &  	$-$02:00:47.2223(1)   & $ 54890 $ \\
J$0645+5158 $ & $     \phn    98.05854629(3)         \phn $ & \phs      $    28.85264422(3)             \phn $  &&	 	06:45:59.081898(9) &  	\phn51:58:14.9208(1)  & $ 56143 $ \\
J$0931-1902 $ & $              152.376967(2) \phn\phn\phn $ &           $     -31.776719(2)     \phn\phn\phn $  &&	 	09:31:19.1180(4)   &   	$-$19:02:55.015(6)    & $ 56469 $ \\
J$1012+5307 $ & $             133.3610921(1)     \phn\phn $ & \phs      $     38.7553210(2)         \phn\phn $  &&	 	10:12:33.43745(6)  &   	\phn53:07:02.3071(7)  & $ 54902 $ \\
J$1024-0719 $ & $            160.73435621(2)         \phn $ &           $   -16.04470826(8)             \phn $  &&	 	10:24:38.670189(9) &   	$-$07:19:19.5396(3)   & $ 55800 $ \\
J$1455-3330 $ & $            231.34753657(5)         \phn $ &           $    -16.0447988(2)         \phn\phn $  &&	 	14:55:47.97069(2)  &   	$-$33:30:46.3833(6)   & $ 55500 $ \\
J$1600-3053 $ & $           244.347677636(9)              $ &           $   -10.07183655(5)             \phn $  &&	 	16:00:51.903261(4) &   	$-$30:53:49.3830(2)   & $ 55416 $ \\
J$1614-2230 $ & $            245.78829040(1)         \phn $ &      \phn $     -1.2567952(5)         \phn\phn $  &&	 	16:14:36.50708(2)  &  	$-$22:30:31.233(2)    & $ 55655 $ \\
J$1640+2224 $ & $            243.98908853(2)         \phn $ & \phs      $    44.05852004(2)             \phn $  &&	 	16:40:16.744825(3) &   	\phn22:24:08.84178(6) & $ 54971 $ \\
J$1643-1224 $ & $            251.08721841(6)         \phn $ & \phs \phn $      9.7783298(4)         \phn\phn $  &&	 	16:43:38.16140(2)  &  	$-$12:24:58.676(1)    & $ 54902 $ \\
J$1713+0747 $ & $           256.668693195(2)              $ & \phs      $   30.700361575(4)                  $  &&	 	17:13:49.5331505(5)&   	\phn07:47:37.49284(2) & $ 54971 $ \\
J$1738+0333 $ & $            264.09490912(9)         \phn $ & \phs      $     26.8842354(1)         \phn\phn $  &&	 	17:38:53.96744(2)  &   	\phn03:33:10.8824(5)  & $ 55800 $ \\
J$1741+1351 $ & $            264.36467815(3)         \phn $ & \phs      $    37.21119890(4)             \phn $  &&	 	17:41:31.144770(5) & 	\phn13:51:44.12241(15)& $ 56176 $ \\
J$1744-1134 $ & $            266.11939556(1)         \phn $ & \phs      $    11.80520366(6)             \phn $  &&	 	17:44:29.407190(3) &  	$-$11:34:54.6925(2)   & $ 54900 $ \\
J$1747-4036 $ & $            267.57913419(5)         \phn $ &           $    -17.2015392(2)         \phn\phn $  &&	 	17:47:48.71665(1)  &    $-$40:36:54.7795(7)   & $ 56281 $ \\
J$1832-0836 $ & $             278.2920105(1)     \phn\phn $ & \phs      $       14.59073(1) \phn\phn\phn\phn $  &&	 	18:32:27.5936(2)   &  	$-$08:36:54.98(4)     & $ 56475 $ \\
J$1853+1303 $ & $            286.25730609(3)         \phn $ & \phs      $    35.74335172(8)             \phn $  &&	 	18:53:57.318423(7) &   	\phn13:03:44.0596(3)  & $ 56155 $ \\
B$1855+09   $ & $            286.86348933(2)         \phn $ & \phs      $    32.32148776(3)             \phn $  &&	 	18:57:36.390614(4) &  	\phn09:43:17.2075(1)  & $ 54978 $ \\
J$1903+0327 $ & $             287.5625804(1)     \phn\phn $ & \phs      $     25.9379873(3)         \phn\phn $  &&	 	19:03:05.79287(3)  &  	\phn03:27:19.194(1)   & $ 55712 $ \\
J$1909-3744 $ & $           284.220863589(4)              $ &           $   -15.15549085(2)             \phn $  &&	 	19:09:47.434674(1) &  	$-$37:44:14.46667(7)  & $ 54500 $ \\
J$1910+1256 $ & $            291.04141433(5)         \phn $ & \phs      $    35.10722400(7)             \phn $  &&	 	19:10:09.70147(1)  &   	\phn12:56:25.4727(2)  & $ 55741 $ \\
J$1918-0642 $ & $            290.31464011(2)         \phn $ & \phs      $    15.35106344(7)             \phn $  &&	 	19:18:48.033256(5) &  	$-$06:42:34.8877(3)   & $ 54901 $ \\
J$1923+2515 $ & $            297.98095593(9)         \phn $ & \phs      $     46.6962061(1)         \phn\phn $  &&	 	19:23:22.49331(2)  &  	\phn25:15:40.6164(5)  & $ 56100 $ \\
B$1937+21   $ & $            301.97324443(1)         \phn $ & \phs      $    42.29675249(1)             \phn $  &&	 	19:39:38.561227(2) &   	\phn21:34:59.12567(5) & $ 54931 $ \\
J$1944+0907 $ & $            299.99545059(3)         \phn $ & \phs      $    29.89102681(8)             \phn $  &&	 	19:44:09.329903(7) &  	\phn09:07:23.0362(3)  & $ 56176 $ \\
J$1949+3106 $ & $              308.657405(2) \phn\phn\phn $ & \phs      $      50.930913(2)     \phn\phn\phn $  &&	 	19:49:29.6379(4)   &   	\phn31:06:03.795(5)   & $ 56367 $ \\
B$1953+29   $ & $            309.6913457(3)      \phn\phn $ & \phs      $     48.6845469(2)         \phn\phn $  &&	 	19:55:27.87546(5)  &   	\phn29:08:43.4464(5)  & $ 56176 $ \\
J$2010-1323 $ & $            301.92448717(2)         \phn $ & \phs \phn $      6.4909501(1)         \phn\phn $  &&	 	20:10:45.920937(8) &   	$-$13:23:56.0755(5)   & $ 55657 $ \\
J$2017+0603 $ & $            308.26117978(7)         \phn $ & \phs      $     25.0444945(1)         \phn\phn $  &&	 	20:17:22.70503(1)  &   	\phn06:03:05.5686(4)  & $ 56200 $ \\
J$2043+1711 $ & $            318.86848854(1)         \phn $ & \phs      $    33.96432619(3)             \phn $  &&	 	20:43:20.882230(3) &   	\phn17:11:28.92694(9) & $ 56175 $ \\
J$2145-0750 $ & $            326.02462112(3)         \phn $ & \phs \phn $      5.3130554(3)         \phn\phn $  &&	 	21:45:50.46089(3)  &   	$-$07:50:18.491(1)    & $ 54903 $ \\
J$2214+3000 $ & $             348.8091355(8)     \phn\phn $ & \phs      $     37.7131533(9)         \phn\phn $  &&	 	22:14:38.85097(2)  &   	\phn30:00:38.1976(2)  & $ 56222 $ \\
J$2302+4442 $ & $ \phn\phn      9.7804392(2)     \phn\phn $ & \phs      $    45.66543639(7)             \phn $  &&	 	23:02:46.97878(3)  &   	\phn44:42:22.0928(3)  & $ 56279 $ \\
J$2317+1439 $ & $            356.12940547(2)         \phn $ & \phs      $    17.68023064(7)             \phn $  &&	 	23:17:09.236644(9) &   	\phn14:39:31.2557(2)  & $ 54977 $ 
\enddata
\tablenotetext{a}{Numbers in parentheses are uncertainties in last digits quoted.}
\tablenotetext{b}{Epoch of position is an exact integer MJD; e.g., 56719 means 56719.000000.}
\end{deluxetable*}

\clearpage

\makeatletter{}\begin{deluxetable*}{lcccccccccc}
\centering
\tablewidth{0pt}
\tabletypesize{\footnotesize}
\tablecaption{Proper Motions\tablenotemark{a}\label{tab:propmo}}
\tablehead{\colhead{PSR} & \colhead{Span} & \colhead{$\mu_\lambda = \dot{\lambda} \mbox{ cos }\beta$} & \colhead{$\mu_\beta = \dot{\beta}$} & \colhead{$\mu_\alpha = \dot{\alpha} \mbox{ cos }\delta$} & \colhead{$\mu_\delta = \dot{\delta}$} & \colhead{$\mu_l = \dot{l} \mbox{ cos }b$} & \colhead{$\mu_b = \dot{b}$}  & \multicolumn{3}{c}{Best Previous Measurement}\\[1pt] \cline{9-11}
& \colhead{(y)} & \colhead{(mas yr$^{-1}$)} & \colhead{(mas yr$^{-1}$)} & \colhead{(mas yr$^{-1}$)} & \colhead{(mas yr$^{-1}$)} & \colhead{(mas yr$^{-1}$)} & \colhead{(mas yr$^{-1}$)}  & \colhead{$\mu_\alpha = \dot{\alpha} \mbox{ cos }\delta$} & \colhead{$\mu_\delta = \dot{\delta}$}& \colhead{Reference}\\&&&&&&& & \colhead{(mas yr$^{-1}$)} & \colhead{(mas yr$^{-1}$)} & \\
}

\startdata
J$0023+0923$ & 2.3 &           $-13.9(2) \phn\phn $  &       \phn $-1\phd(1)\phn\phn\phn $ &            $ -12.3(6)       \phn\phn\phn     $  & \phn       $  -6.7(9)    \phn\phn     $  &          $-13.19$ &     \phn $      -4.60 $   &                  \nodata                 & \nodata & \nodata \\ 
J$0030+0451$ & 8.8 &      \phn $-5.52(1) \phn     $  &  \phs \phn $3.0(5) \phn\phn     $ &       \phn $  -6.3(2)       \phn\phn\phn     $  & \phs  \phn $  0.6(5)     \phn\phn     $  & \phn     $-6.13 $ & \phs\phn $      1.42 $    &      \phn \phn $ -5.3(9)    \phn\phn    \phn $ &      \phn \phn $ -2\phd(2)  \phn\phn\phn\phn $ & 1\phantom{$^*$} \\ 
J$0340+4130$ & 1.7 &      \phn $-2.4(8)  \phn\phn $  &       \phn $-4\phd(1)\phn\phn\phn $ &       \phn $  -1.3(7)       \phn\phn\phn     $  &       \phn $  -5\phd(1)     \phn\phn\phn $  & \phs\phn $1.86  $ &     \phn $      -4.52$    &                  \nodata                 & \nodata & \nodata \\ 
J$0613-0200$ & 8.6 & \phs \phn $2.12(2)  \phn     $  &            $-10.34(4) \phn      $ & \phs  \phn $   1.85(2)      \phn\phn         $  &            $  -10.39(4)  \phn         $  & \phs     $10.08 $ &     \phn $      -3.15$    &  \phs \phn \phn $ 1.84(4)   \phn        \phn $ &          \phn $ -10.6(1)    \phn\phn    \phn $ & 2\phantom{$^*$} \\ 
J$0645+5158$ & 2.4 & \phs \phn $2.1(1)    \phn\phn $  &       \phn $-7.3(2) \phn\phn   $ & \phs  \phn $   1.4(1)       \phn\phn\phn     $  &       \phn $  -7.5(2)    \phn\phn     $  & \phs\phn $7.52  $ &     \phn $      -0.98$    &  \phs \phn \phn $ 1.2(1)    \phn\phn    \phn $ &      \phn \phn $ -7.5(2)    \phn\phn    \phn $ & 3\phantom{$^*$} \\ 
J$0931-1902$ & 0.6 & \nodata                          &            \nodata               &           \nodata                               &             \nodata                      &          \nodata  & \nodata                   &                  \nodata                 & \nodata & \nodata \\ 
J$1012+5307$ & 9.2 & \phs      $13.9(1)  \phn\phn $  &            $-21.7(3) \phn\phn   $ & \phs  \phn $   2.5(2)       \phn\phn\phn     $  &            $  -25.6(2)   \phn\phn     $  & \phs     $21.89 $ & \phs     $      13.57 $   &  \phs \phn \phn $ 2.562(14)                  $ &          \phn $ -25.61(2)   \phn        \phn $ & 4\phantom{$^*$} \\ 
J$1024-0719$ & 4.0 &           $-14.36(6)\phn     $  &            $-57.8(3) \phn\phn   $ &            $ -35.2(1)       \phn\phn\phn     $  &            $  -48.0(2)   \phn\phn     $  & \phs\phn $7.73  $ &          $      -59.03$   &          \phn $ -35.3(1)    \phn\phn    \phn $ &          \phn $ -48.2(2)    \phn\phn    \phn $ & 2\phantom{$^*$} \\ 
J$1455-3330$ & 9.2 & \phs \phn $8.16(7)  \phn     $  &  \phs \phn $0.5(3) \phn\phn     $ & \phs  \phn $   7.9(1)       \phn\phn\phn     $  &       \phn $  -2.0(3)    \phn\phn     $  & \phs\phn $5.85  $ &     \phn $      -5.71 $   &  \phs \phn \phn $ 5\phd(6)     \phn\phn\phn\phn $ &  \phs     \phn $ 24\phd(12) \phn\phn\phn     $ & 5\phantom{$^*$} \\ 
J$1600-3053$ & 6.0 & \phs \phn $0.47(2)  \phn     $  &       \phn $-7.0(1)    \phn\phn $ &       \phn $  -0.95(3)      \phn\phn         $  &       \phn $  -7.0(1)    \phn\phn     $  &     \phn $-5.47 $ &     \phn $      -4.42 $   &      \phn \phn $ -1.06(5)   \phn        \phn $ &      \phn \phn $ -7.1(2)    \phn\phn    \phn $ & 2\phantom{$^*$} \\ 
J$1614-2230$ & 5.1 & \phs \phn $9.46(2)  \phn     $  &            $-31\phd(1)\phn\phn\phn $ & \phs  \phn $   3.8(2)       \phn\phn\phn     $  &            $  -32\phd(1)    \phn\phn\phn $  &          $-21.19$ &          $      -24.65$   &                  \nodata                 & \nodata & \nodata \\ 
J$1640+2224$ & 8.9 & \phs \phn $4.20(1)  \phn     $  &            $-10.73(2) \phn      $ & \phs  \phn $   2.09(1)      \phn\phn         $  &            $  -11.33(2)  \phn         $  &          $-10.12$ &     \phn $      -5.5  $   &  \phs \phn \phn $ 2.10(3)   \phn        \phn $ &          \phn $ -11.20(7)   \phn        \phn $ & 6\phantom{$^*$} \\ 
J$1643-1224$ & 9.0 & \phs \phn $5.56(8)  \phn     $  &  \phs \phn $5.3(5) \phn\phn     $ & \phs  \phn $   6.2(1)       \phn\phn\phn     $  & \phs  \phn $  4.5(5)     \phn\phn     $  & \phs\phn $7.27  $ &     \phn $      -2.39 $   &  \phs \phn \phn $ 5.99(5)   \phn        \phn $ &  \phs \phn \phn $ 4.1(2)    \phn\phn    \phn $ & 2\phantom{$^*$} \\ 
J$1713+0747$ & 8.8 & \phs \phn $5.260(2)          $  &       \phn $-3.442(5)           $ & \phs  \phn $   4.918(2)     \phn             $  &       \phn $  -3.914(5)               $  &     \phn $-1.29 $ &     \phn $      -6.15 $   &  \phs \phn \phn $ 4.915(3)              \phn $ &      \phn \phn $ -3.914(5)              \phn $ & 7$^*$ \\ 
J$1738+0333$ & 4.0 & \phs \phn $6.6(2)    \phn\phn $  &  \phs \phn $6.0(4) \phn\phn    $ & \phs  \phn $   6.9(2)       \phn\phn\phn     $  & \phs  \phn $  5.8(4)     \phn\phn     $  & \phs\phn $8.28  $ &     \phn $      -3.43 $   &  \phs \phn \phn $ 7.037(5)              \phn $ &  \phs \phn \phn $ 5.073(12)                  $ & 8\phantom{$^*$} \\ 
J$1741+1351$ & 2.3 &      \phn $-8.8(1) \phn\phn  $  &       \phn $-7.6(2) \phn\phn    $ &       \phn $  -9.1(1)       \phn\phn\phn     $  &       \phn $  -7.2(2)    \phn\phn     $  &          $-10.35$ & \phs\phn $      5.27  $   &                  \nodata                 & \nodata & \nodata \\ 
J$1744-1134$ & 9.2 & \phs      $19.01(2) \phn     $  &       \phn $-8.68(8) \phn       $ & \phs       $  18.76(2)      \phn\phn         $  &       \phn $  -9.20(8)   \phn         $  & \phs\phn $1.56  $ &          $      -20.84$   &  \phs     \phn $ 18.804(8)              \phn $ &      \phn \phn $ -9.40(3)   \phn        \phn $ & 2\phantom{$^*$} \\ 
J$1747-4036$ & 1.7 & \phs \phn $0.1(8)   \phn\phn $  &       \phn $-6\phd(1)\phn\phn\phn $ & \phs  \phn $   0\phd(1)   \phn\phn\phn\phn $  &       \phn $  -6\phd(1)  \phn\phn\phn $  &     \phn $-5.10 $ &     \phn $      -2.96 $   &                  \nodata                 & \nodata & \nodata \\ 
J$1832-0836$ & 0.6 &  \nodata                         &            \nodata               &               \nodata                           &               \nodata                    &           \nodata & \nodata                   &                  \nodata                 & \nodata & \nodata \\ 
J$1853+1303$ & 2.3 &      \phn $-1.82(15)         $  &       \phn $-2.9(4) \phn\phn    $ &       \phn $  -1.48(2)      \phn\phn         $  &       \phn $  -3.1(4)    \phn\phn     $  &     \phn $-3.40 $ &     \phn $      -0.07 $   &      \phn \phn $ -1.68(4)   \phn        \phn $ &      \phn \phn $ -2.94(6)   \phn        \phn $ & 9$^*$ \\ 
B$1855+09  $ & 8.9 &      \phn $-3.27(1) \phn     $  &       \phn $-5.10(3) \phn       $ &       \phn $  -2.651(15)                     $  &       \phn $  -5.45(3)   \phn         $  &     \phn $-6.06 $ &     \phn $      -0.15 $   &      \phn \phn $ -2.64(2)   \phn        \phn $ &      \phn \phn $ -5.46(2)   \phn        \phn $ & 2\phantom{$^*$} \\ 
J$1903+0327$ & 4.0 &      \phn $-3.5(3)  \phn\phn $  &       \phn $-6.2(9) \phn\phn    $ &       \phn $  -2.7(3)       \phn\phn\phn     $  &       \phn $  -6.5(9)    \phn\phn     $  &     \phn $-7.07 $ &     \phn $      -0.59 $   &      \phn \phn $ -2.06(7)   \phn        \phn $ &      \phn \phn $ -5.21(12)  \phn             $ & 10\phantom{$^*$} \\ 
J$1909-3744$ & 9.1 &           $-13.868(4)        $  &            $-34.34(2) \phn      $ &       \phn $  -9.518(4)     \phn             $  &            $  -35.79(2)  \phn         $  &          $-36.91$ &     \phn $      -3.04 $   &      \phn \phn $ -9.510(4)              \phn $ &          \phn $ -35.859(10)                  $ & 2\phantom{$^*$} \\ 
J$1910+1256$ & 4.7 &      \phn $-0.7(1) \phn\phn  $  &       \phn $-7.2(2) \phn\phn    $ & \phs  \phn $   0.3(1)       \phn\phn\phn     $  &       \phn $  -7.2(2)    \phn\phn     $  &     \phn $-6.20 $ &     \phn $      -3.66 $   &  \phs \phn \phn $ 0.21(5)   \phn        \phn $ &      \phn \phn $ -7.25(6)   \phn        \phn $ & 9$^*$ \\ 
J$1918-0642$ & 9.0 &      \phn $-7.93(2) \phn     $  &       \phn $-4.85(9) \phn       $ &       \phn $  -7.18(3)      \phn\phn         $  &       \phn $  -5.90(9)   \phn         $  &     \phn $-8.50 $ & \phs\phn $      3.75  $   &      \phn \phn $ -7.20(10)  \phn             $ &      \phn \phn $ -5.7(3)    \phn\phn    \phn $ & 11\phantom{$^*$} \\ 
J$1923+2515$ & 2.2 &      \phn $-9.5(2)  \phn\phn $  &            $-12.8(5) \phn\phn   $ &       \phn $  -6.6(2)       \phn\phn\phn     $  &       \phn $  -14.5(5)   \phn\phn     $  &          $-15.88$ &     \phn $      -1.04 $   &      \phn \phn $ -6.2(24)   \phn\phn         $ &          \phn $ -23.5(70)   \phn\phn         $ & 12\phantom{$^*$} \\ 
B$1937+21  $ & 9.1 &      \phn $-0.02(1) \phn     $  &       \phn $-0.41(2) \phn       $ & \phs  \phn $   0.07(1)      \phn\phn         $  &       \phn $  -0.40(2)   \phn         $  &     \phn $-0.31 $ &     \phn $      -0.27 $   &  \phs \phn \phn $ 0.072(1)              \phn $ &      \phn \phn $ -0.415(2)              \phn $ & 2\phantom{$^*$} \\ 
J$1944+0907$ & 2.3 & \phs \phn $9.42(13)          $  &            $-25.5(4) \phn\phn   $ & \phs       $  14.37(11)     \phn             $  &            $  -23.1(4)   \phn\phn     $  &          $-13.03$ &          $      -23.88$   &  \phs     \phn $ 12.0(7)    \phn\phn    \phn $ &          \phn $ -18\phd(3)  \phn\phn\phn\phn $ & 13\phantom{$^*$} \\ 
J$1949+3106$ & 1.2 & \phs      $13\phd(15)  \phn\phn $  &  \phs   $10\phd(13)  \phn\phn$ & \phs       $  10\phd(11)    \phn\phn\phn     $  & \phs       $  13\phd(16) \phn\phn     $  & \phs     $16.50 $ &     \phn $      -1.79 $   &      \phn \phn $ -2.94(6)   \phn        \phn $ &      \phn \phn $ -5.17(8)   \phn        \phn $ & 14\phantom{$^*$} \\ 
B$1953+29  $ & 2.3 &      \phn $-1.8(9) \phn\phn  $  &       \phn $-4.4(14) \phn       $ &       \phn $  -0.4(12)      \phn\phn\phn     $  &       \phn $  -5\phd(1)  \phn\phn\phn $  &     \phn $-4.24 $ &     \phn $      -2.11 $   &      \phn \phn $ -0.9(1)    \phn\phn    \phn $ &      \phn \phn $ -4.1(1)    \phn\phn    \phn $ & 9$^*$ \\ 
J$2010-1323$ & 4.1 & \phs \phn $1.16(4)  \phn     $  &       \phn $-7.3(4)  \phn\phn   $ & \phs  \phn $   2.71(9)      \phn\phn         $  &       \phn $  -6.9(4)    \phn\phn     $  &     \phn $-5.13 $ &     \phn $      -5.29 $   &                  \nodata                 & \nodata & \nodata \\ 
J$2017+0603$ & 1.7 & \phs \phn $2.3(6)   \phn\phn $  &       \phn $-0.1(7)  \phn\phn   $ & \phs  \phn $   2.2(7)       \phn\phn\phn     $  & \phs  \phn $  0.5(6)     \phn\phn     $  & \phs\phn $1.55  $ &     \phn $      -1.7  $   &                  \nodata                 & \nodata & \nodata \\ 
J$2043+1711$ & 2.3 &      \phn $-8.97(7) \phn     $  &       \phn $-8.5(1)  \phn\phn   $ &       \phn $  -5.85(7)      \phn\phn         $  &            $  -10.9(1)   \phn\phn     $  &          $-12.26$ &     \phn $      -1.67 $   &      \phn \phn $ -7\phd(2   \phn\phn\phn\phn $ &          \phn $ -11\phd(2)  \phn\phn\phn\phn $ & 15\phantom{$^*$} \\ 
J$2145-0750$ & 9.1 &           $-12.04(4)\phn     $  &       \phn $-3.7(4)  \phn\phn   $ &            $ -10.1(1)       \phn\phn\phn     $  &       \phn $  -7.5(4)    \phn\phn     $  &          $-11.55$ & \phs\phn $      4.90  $   &      \phn \phn $ -9.66(8)   \phn        \phn $ &      \phn \phn $ -8.9(2)    \phn\phn    \phn $ & 2\phantom{$^*$} \\ 
J$2214+3000$ & 2.1 & \phs      $17.1(5)  \phn\phn $  &            $-10.5(9) \phn\phn   $ & \phs       $  20.0(6)       \phn\phn\phn     $  &       \phn $  -1.7(8)    \phn\phn     $  & \phs     $15.07 $ &          $      -13.25$   &                  \nodata                 & \nodata & \nodata \\ 
J$2302+4442$ & 1.7 &      \phn $-3.3(6)  \phn\phn $  &       \phn $-1\phd(2) \phn\phn\phn $ &       \phn $  -2\phd(1)  \phn\phn\phn\phn $  &       \phn $  -3\phd(2)     \phn\phn\phn $  &     \phn $-2.97 $ &     \phn $      -1.98 $   &                  \nodata                 & \nodata & \nodata \\ 
J$2317+1439$ & 8.9 & \phs \phn $0.19(2)  \phn     $  &  \phs \phn $3.80(7) \phn        $ &       \phn $  -1.39(3)      \phn\phn         $  & \phs  \phn $  3.55(6)    \phn         $  & \phs\phn $0.44  $ & \phs\phn $      3.78  $   &      \phn \phn $ -1.7(15)   \phn\phn         $ &  \phs \phn \phn $ 7.4(31)   \phn\phn         $ & 16\phantom{$^*$}  
\enddata
\tablecomments{Numbers in parentheses are 1$\sigma$ uncertainties in the last digit quoted.
In references 2 and 9, 2$\sigma$ uncertainties were reported; we quote half those uncertainties here except in circumstances when the uncertainty digit was reported as 1.)
{\bf References:}
(1) \cite{Abdo2009},
(2) \cite{Verbiest2009},
(3) \cite{Stovall2014},
(4) \cite{Lazaridis2009},
(5) \cite{Toscano1999},
(6) \cite{Hou2014},
(7) \cite{Zhu2015},
(8) \cite{Freire2012},
(9) \cite{Gonzalez2011},
(10) \cite{Freire2011},
(11) \cite{Janssen2010},
(12) \cite{Lynch2013},
(13) \cite{Champion2005},
(14) \cite{Deneva2012},
(15) \cite{Guillemot2012},
(16) \cite{Camilo1996},
References marked by asterisks used some of the same data as the present work.}
\end{deluxetable*}

\clearpage

\makeatletter{}\begin{deluxetable}{lccc}
\centering
\tablewidth{0pt}
\tabletypesize{\small}
\tablecaption{Parallaxes\label{tab:px}}
\tablehead{\colhead{PSR} & \colhead{Parallax} & \multicolumn{2}{c}{Best Previous Measurement} \\[1pt]
\cline{3-4}
 & \colhead{(mas)} & \colhead{Parallax} & \colhead{Reference \rule{0pt}{10pt}}
\\
 & & (mas) & 
}
\startdata
J$0023+0923$ & \phs $  0.4(3) \phn     $  & \nodata                           & \nodata \\   
J$0030+0451$ & \phs $  3.3(2) \phn     $  & $\phantom{<} 3.3(5) \phn $        & 1 \\ 
J$0340+4130$ & \phs $  0.7(7) \phn     $  & \nodata                           & \nodata \\  
J$0613-0200$ & \phs $  0.9(2) \phn     $  & $\phantom{<} 0.8(4) \phn $        & 2 \\  
J$0645+5158$ & \phs $  1.3(3) \phn     $  & $\phantom{<} 1.4(4) \phn $        & 3 \\ 
J$0931-1902$ & \phs $  8\phd(8)  \phn\phn $  & \nodata                           & \nodata \\  
J$1012+5307$ & \phs $  1\phd(3)  \phn\phn $  & $\phantom{<} 1.2(3) \phn $        & 4\\ 
J$1024-0719$ & \phs $  0.6(3) \phn     $  & $\phantom{<} 1.9(8) \phn $        & 5\\  
J$1455-3330$ & \phs $  0.2(6) \phn     $  & \nodata                           & \nodata \\  
J$1600-3053$ & \phs $  0.34(9)         $  & $\phantom{<} 0.2(2) \phn $        & 2\\  
J$1614-2230$ & \phs $  1.5(1) \phn     $  & $\phantom{<} 1.5(1) \phn $        & 6\\ 
J$1640+2224$ &      $ -1.0(6) \phn     $  & $        {<} 3.7 \phantom{0(0)} $ & 7\\ 
J$1643-1224$ & \phs $  0.7(6) \phn     $  & $\phantom{<} 2.2(4) \phn $        & 2\\  
J$1713+0747$ & \phs $  0.85(3)         $  & $\phantom{<} 0.94(5)$             & 2\\  
J$1738+0333$ & \phs $  0.4(5) \phn     $  & $\phantom{<} 0.68(5)$             & 8,9  \\  
J$1741+1351$ & \phs $  0.0(5) \phn     $  & \nodata                           & \nodata \\  
J$1744-1134$ & \phs $  2.4(1) \phn     $  & $\phantom{<} 2.4(1) \phn $        & 2 \\ 
J$1747-4036$ &      $ -0.4(7) \phn     $  & \nodata                           & \nodata \\ 
J$1832-0836$ & \phs $  5\phd(5)  \phn\phn $  & \nodata                           & \nodata \\   
J$1853+1303$ & \phs $  0.1(5) \phn     $  & $\phantom{<} 1.0(6) \phn $        & 10 \\   
B$1855+09  $ & \phs $  0.3(2) \phn     $  & $\phantom{<} 1.1(1) \phn $        & 2 \\    
J$1903+0327$ & \phs $  0.4(8) \phn     $  & \nodata                           & \nodata \\  
J$1909-3744$ & \phs $  0.94(3)         $  & $\phantom{<} 0.79(2) $            & 2 \\  
J$1910+1256$ &      $ -0.3(7) \phn     $  & $        {<} 0.7 \phantom{0(0)} $ & 10 \\  
J$1918-0642$ & \phs $  1.1(2) \phn     $  & \nodata                           & \nodata \\   
J$1923+2515$ & \phs $  2\phd(1)  \phn\phn $  & \nodata                           & \nodata \\ 
B$1937+21  $ & \phs $  0.1(1) \phn     $  & \nodata                           & \nodata \\    
J$1944+0907$ & \phs $  0.0(4) \phn     $  & \nodata                           & \nodata \\  
J$1949+3106$ &      $ -6\phd(7)  \phn\phn $  & \nodata                           & \nodata \\
B$1953+29  $ &      $ -4\phd(2)  \phn\phn $  & \nodata                           & \nodata \\  
J$2010-1323$ & \phs $  0.1(2) \phn     $  & \nodata                           & \nodata \\  
J$2017+0603$ & \phs $  0.4(3) \phn     $  & \nodata                           & \nodata \\  
J$2043+1711$ & \phs $  0.8(2) \phn     $  & \nodata                           & \nodata \\  
J$2145-0750$ & \phs $  1.3(2) \phn     $  & \phs $1.6(3) \phn $               & 2\\ 
J$2214+3000$ & \phs $  1\phd(1)  \phn\phn $  & \nodata                           & \nodata \\  
J$2302+4442$ &      $ -2\phd(2)  \phn\phn $  & \nodata                           & \nodata \\
J$2317+1439$ & \phs $  0.7(2) \phn     $  & \nodata                           & \nodata 
\enddata
\tablecomments{Numbers in parentheses are 1$\sigma$ uncertainties in the last digit quoted.
(In references 1, 2, and 10, 2$\sigma$ uncertainties were reported; we quote
half those uncertainties here.)
{\bf References:} 
(1) \cite{Lommen2006}; 
(2) \cite{Verbiest2009}; 
(3) \cite{Stovall2014}; 
(4) \cite{Lazaridis2009}; 
(5) \cite{Hotan2006}; 
(6) \cite{Fermi2013};
(7) \cite{Lohmer2005}; 
(8) \cite{Antoniadis2012}; 
(9) \cite{Freire2012};
(10) \cite{Gonzalez2011}. 
}
\end{deluxetable}

\clearpage

\makeatletter{}\begin{deluxetable}{lllc}
  \centering
  \tablewidth{0pt}
  \tabletypesize{\small}
  \tablecaption{Distances from Parallax Measurements
  \label{tab:pxlimits}}
  \tablehead{\colhead{PSR} & \colhead{Parallax} & \colhead{Distance} & \colhead{NE2001 Distance}\\
  & \colhead{(mas)} & \colhead{(kpc)} & \colhead{(kpc)}} 
\startdata
\multicolumn{4}{c}{Distance Measurements} \\[2pt]
\tableline
\rule{0pt}{10pt}\vspace{1.40mm}J$0030+0451 $ & $ \phantom{<} 3.3(2) \phn $ & $ \phantom{>} 0.30^{+0.02}_{-0.01}     $ & $0.32$\\ 
\vspace{1.40mm}J$0613-0200 $ & $ \phantom{<} 0.9(2) \phn $ & $ \phantom{>} 1.1^{+0.2}_{-0.2}   \phn $ & $1.70$\\
\vspace{1.40mm}J$0645+5158 $ & $ \phantom{<} 1.3(3) \phn $ & $ \phantom{>} 0.8^{+0.3}_{-0.2}   \phn $ & $0.70$\\
\vspace{1.40mm}J$1600-3053 $ & $ \phantom{<} 0.34(9)     $ & $ \phantom{>} 3.0^{+1.0}_{-0.6}   \phn $ & $1.63$\\
\vspace{1.40mm}J$1614-2230 $ & $ \phantom{<} 1.5(1) \phn $ & $ \phantom{>} 0.65^{+0.05}_{-0.04}     $ & $1.27$\\
\vspace{1.40mm}J$1713+0747 $ & $ \phantom{<} 0.85(3)\phn $ & $ \phantom{>} 1.18^{+0.04}_{-0.04}     $ & $0.89$\\
\vspace{1.40mm}J$1744-1134 $ & $ \phantom{<} 2.4(1) \phn $ & $ \phantom{>} 0.41^{+0.02}_{-0.02}     $ & $0.42$\\
\vspace{1.40mm}J$1909-3744 $ & $ \phantom{<} 0.94(3)     $ & $ \phantom{>} 1.07^{+0.04}_{-0.03}     $ & $0.46$\\
\vspace{1.40mm}J$1918-0642 $ & $ \phantom{<} 1.1(2) \phn $ & $ \phantom{>} 0.9^{+0.2}_{-0.1}   \phn $ & $1.24$\\
\vspace{1.40mm}J$2043+1711 $ & $ \phantom{<} 0.8(2) \phn $ & $ \phantom{>} 1.3^{+0.4}_{-0.3}   \phn $ & $1.78$\\
\vspace{1.40mm}J$2145-0750 $ & $ \phantom{<} 1.3(2) \phn $ & $ \phantom{>} 0.8^{+0.2}_{-0.1}   \phn $ & $0.57$\\
J$2317+1439 $ & $ \phantom{<} 0.7(2) \phn $ & $ \phantom{>} 1.3^{+0.4}_{-0.2}   \phn $ & $0.96$\\
\cutinhead{Distance Lower Limits}
\vspace{0.70mm}J$0023+0923 $ &  $	{<}1.00 \phantom{(0)} $  &  ${>}1.00 $ & $0.70$\\
\vspace{0.70mm}J$1012+5307 $ &  $	{<}6.08 \phantom{(0)} $  &  ${>}0.16 $ & $0.41$\\
\vspace{0.70mm}J$1024-0719 $ &  $	{<}1.10 \phantom{(0)} $  &  ${>}0.91 $ & $0.39$\\
\vspace{0.70mm}J$1455-3330 $ &  $	{<}1.36 \phantom{(0)} $  &  ${>}0.74 $ & $0.53$\\
\vspace{0.70mm}J$1640+2224 $ &  $	{<}0.69 \phantom{(0)} $  &  ${>}1.45 $ & $1.16$\\
\vspace{0.70mm}J$1643-1224 $ &  $	{<}1.67 \phantom{(0)} $  &  ${>}0.60 $ & $2.41$\\
\vspace{0.70mm}J$1738+0333 $ &  $	{<}1.22 \phantom{(0)} $  &  ${>}0.82 $ & $1.43$\\
\vspace{0.70mm}J$1741+1351 $ &  $	{<}0.98 \phantom{(0)} $  &  ${>}1.02 $ & $0.90$\\
\vspace{0.70mm}J$1853+1303 $ &  $	{<}1.02 \phantom{(0)} $  &  ${>}0.98 $ & $2.09$\\
\vspace{0.70mm}B$1855+09   $ &  $      {<}0.66 \phantom{(0)} $  &  ${>}1.52 $ & $1.17$\\
\vspace{0.70mm}J$1903+0327 $ &  $	{<}1.86 \phantom{(0)} $  &  ${>}0.54 $ & $6.36$\\
\vspace{0.70mm}J$1910+1256 $ &  $	{<}1.10 \phantom{(0)} $  &  ${>}0.91 $ & $2.33$\\
\vspace{0.70mm}J$1923+2515 $ &  $	{<}4.68 \phantom{(0)} $  &  ${>}0.21 $ & $1.63$\\
\vspace{0.70mm}B$1937+21   $ &  $      {<}0.31 \phantom{(0)} $  &  ${>}3.23 $ & $3.56$\\
\vspace{0.70mm}J$1944+0907 $ &  $	{<}0.74 \phantom{(0)} $  &  ${>}1.36 $ & $1.81$\\
\vspace{0.70mm}B$1953+29   $ &  $      {<}2.71 \phantom{(0)} $  &  ${>}0.37 $ & $4.64$\\
\vspace{0.70mm}J$2010-1323 $ &  $	{<}0.43 \phantom{(0)} $  &  ${>}2.33 $ & $1.03$\\
J$2214+3000 $ &  $	{<}2.86 \phantom{(0)} $  &  ${>}0.35 $ & $1.58$
  \enddata
\end{deluxetable}

\clearpage

\makeatletter{}\begin{deluxetable}{lccc}
  \centering
  \tablewidth{0pt}
  \tabletypesize{\small}
  \tablecaption{Distances from $\dot{P}_{b}$ Measurements
  \label{tab:pbdotdistances}
  }
  \tablehead{\colhead{PSR} & \colhead{$P_{\rm b}$} & \colhead{$\dot{P}_{\rm b}$} & \colhead{$d_{\dot{P}_{\rm b}}$} \\
  & \colhead{(days)} & \colhead{($10^{-12}$)} & \colhead{(kpc)}} 
\startdata
\multicolumn{4}{c}{Distance Measurements} \\[2pt]
\tableline
\rule{0pt}{10pt}J$1614-2230$  & \phn\phn  8.687   & $\phantom{-}\phn 1.3(7)   \phn\phn $   & $ \phantom{<} 0.7(3)  \phn    $ \\
J$1909-3744$  & \phn\phn  1.533   & $\phantom{-}\phn 0.506(8)          $   & $ \phantom{<} 1.11(2)         $ \\
\cutinhead{Distance Upper Limits}
J$0613-0200$  & \phn\phn  1.199   & \phn  $-0.00(3)   \phn   $       & $ {<} \phn 1.9  \phantom{(0)} $ \\
J$1012+5307$  & \phn\phn  0.604   & \phn  $-0.03(15)         $       & $ {<} \phn 3.9  \phantom{(0)} $ \\
J$1640+2224$  &          175.461  &       $-15\phd(22) \phn\phn $       & $ {<}     10.2  \phantom{(0)} $ \\
J$1918-0642$  & \phn      10.913  & \phn  $-0.5(5)  \phn\phn $       & $ {<}     10.1  \phantom{(0)} $ \\
J$2145-0750$  & \phn\phn   6.839  & \phn  $-0.2(2)  \phn\phn $       & $ {<} \phn 2.4  \phantom{(0)} $ 
\enddata
\end{deluxetable}

\clearpage

\makeatletter{}\begin{deluxetable}{lccc}
  \centering
  \tablewidth{0pt}
  \tabletypesize{\small}
  \tablecaption{Distance Upper Limits from $\dot{P}$ Measurements
  \label{tab:pdotdistances}
  }
  \tablehead{\colhead{PSR} & \colhead{$P$} & \colhead{$\dot{P}$} & \colhead{$d_{\dot{P}}$} \\
  & \colhead{(ms)} & \colhead{($10^{-21}$)} & \colhead{(kpc)}} 
\startdata
J0023+0923 &  \phn  3.050 &          11.421 & $ {<}\phn  7.1$ \\
J0030+0451 &  \phn  4.865 &          10.174 & $ {<}     13.6$ \\
J0613-0200 &  \phn  3.062 &   \phn    9.590 & $ {<}     10.3$ \\
J0645+5158 &  \phn  8.853 &   \phn    4.920 & $ {<}\phn  3.3$ \\
J1012+5307 &  \phn  5.256 &          17.127 & $ {<}\phn  2.1$ \\
J1024-0719 &  \phn  5.162 &          18.552 & $ {<}\phn  0.4$ \\
J1614-2230 &  \phn  3.151 &   \phn    9.624 & $ {<}\phn  1.2$ \\
J1640+2224 &  \phn  3.163 &   \phn    2.818 & $ {<}\phn  3.4$ \\
J1744-1134 &  \phn  4.075 &   \phn    8.934 & $ {<}\phn  1.9$ \\
J1909-3744 &  \phn  2.947 &          14.025 & $ {<}\phn  1.4$ \\
J1923+2515 &  \phn  3.788 &   \phn    9.553 & $ {<}\phn  4.8$ \\
J1944+0907 &  \phn  5.185 &          17.339 & $ {<}\phn  2.0$ \\
J2010-1323 &  \phn  5.223 &   \phn    4.824 & $ {<}     14.6$ \\
J2043+1711 &  \phn  2.380 &   \phn    5.243 & $ {<}\phn  7.5$ \\
J2145-0750 &       16.052 &          29.790 & $ {<}\phn  4.6$ \\
J2214+3000 &  \phn  3.119 &          14.701 & $ {<}\phn  5.0$ \\
J2317+1439 &  \phn  3.445 &   \phn    2.430 & $ {<}     12.6$ 
\enddata
\end{deluxetable}

\clearpage

\makeatletter{}\begin{deluxetable*}{lccccc}
\centering
\tablewidth{0pt}
\tablecaption{Non-NANOGrav Pulsar Proper Motion Measurements Used in our Analysis\label{tab:propmo_nonNANOGrav}}
\tablehead{\colhead{Pulsar} & \colhead{$\mu_{\alpha}=\dot{\alpha}\mbox{ cos}\delta$} & \colhead{$\mu_{\delta}=\dot{\delta}$} & \colhead{$\mu_l = \dot{l}\mbox{ cos}b$} & \colhead{$\mu_b = \dot{b}$} & \colhead{Reference} \\ & \colhead{(mas/year)} & \colhead{(mas/year)} & \colhead{(mas/year)} & \colhead{(mas/year)} & }

\startdata
J$0101-6422$  &	\phs \phn       $10\phd(1)     \phn\phn\phn\phn  $ &           $-12\phd(2) \phn\phn\phn\phn$ &      \phn $-9.03	$ & \phs \phn      $12.75 $ &  $1$\\
J$0218+4232$  &	\phs \phn \phn  $5.35(5)    \phn\phn          $ &      \phn $-3.74(12) \phn          $ & \phs \phn $6.32	$ &      \phn \phn $-1.62 $ &  $2$\\
J$0437-4715$  &	\phs            $121.679(52)                  $ &           $-71.820(86)             $ & \phs      $64.84	$ & \phs           $125.54$ &  $3$\\
J$0610-2100$  &	\phs \phn \phn  $9.0(2)     \phn\phn\phn\phn  $ & \phs      $17.1(2)    \phn\phn\phn $ &           $-12.16	$ & \phs \phn      $15.02 $ &  $4$\\
J$0711-6830$  &	     \phn       $-15.55(8)  \phn\phn          $ & \phs      $14.23(7) \phn\phn       $ &           $-17.94	$ &      \phn      $-11.07$ &  $5$\\ 
J$0751+1807$  &      \phn \phn  $-1.3(2)    \phn\phn\phn      $ &      \phn $-6\phd(2) \phn\phn\phn\phn $ & \phs \phn $5.01        $ &      \phn \phn $-3.54 $ &  $6$\\
J$1012+5307$  &	\phs \phn \phn  $2.562(14)                    $ &           $-25.61(2) \phn\phn      $ & \phs      $21.86	$ & \phs \phn      $13.58 $ &  $7$\\ 
J$1017-7156$  &	     \phn \phn  $-7.31(6)   \phn\phn          $ & \phs \phn $6.76(5) \phn\phn        $ &      \phn $-9.84	$ & \phs \phn \phn $1.53  $ &  $8$\\ 
J$1023+0038$  &	\phs \phn \phn  $4.76(3)    \phn\phn          $ &           $-17.34(4) \phn\phn      $ & \phs      $16.25	$ &      \phn \phn $-7.70  $ &  $9$\\ 
J$1045-4509$  &	     \phn \phn  $-6.0(2)    \phn\phn\phn      $ & \phs \phn $5.3(2) \phn\phn\phn     $ &      \phn $-7.78	$ & \phs \phn \phn $1.90   $ &  $5$\\
J$1125-5825$  &	     \phn       $-10.0(3)   \phn\phn\phn      $ & \phs \phn $2.4(3) \phn\phn\phn     $ &           $-10.25	$ &      \phn \phn $-0.88 $ &  $8$\\
J$1231-1411$  &	          \phn  $-60\phd(4)  \phn\phn\phn\phn $ &       $14\phd(8)   \phn\phn\phn\phn$ &           $-61.66	$ &      \phn \phn $7.95  $ &  $4$\\ 
B$1257+12  $  & \phs \phn       $45.50(5)   \phn\phn          $ &           $-84.70(7) \phn\phn      $ & \phs      $30.80	$ &      \phn      $-91.08$ &  $10$\\   
J$1446-4701$  &	     \phn \phn  $-4.0(2)    \phn\phn\phn      $ &      \phn $-2.0(3) \phn\phn\phn    $ &      \phn $-4.47	$ &      \phn \phn $-0.02 $ &  $8$\\ 
J$1603-7202$  &      \phn \phn  $-2.52(6)   \phn\phn          $ &      \phn $-7.42(9) \phn\phn       $ &      \phn $-6.94       $ &      \phn \phn $-3.64 $ &  $5$\\
J$1738+0333$  &	\phs \phn \phn  $7.037(5)   \phn              $ & \phs \phn $5.073(12)               $ & \phs \phn $7.76	$ &      \phn \phn $-3.88 $ &  $11$\\  
J$1745+1017$  &	\phs \phn \phn  $6\phd(1)	    \phn\phn\phn\phn  $ &      \phn $-5\phd(1) \phn\phn\phn\phn $ &      \phn $-1.91	$ &      \phn \phn $-7.57 $ &  $12$\\  
J$1843-1113$  &	     \phn \phn  $-2.17(7)   \phn\phn          $ &      \phn $-2.74(25) \phn          $ &      \phn $-3.43	$ & \phs \phn \phn $0.68  $ &  $13$\\    
J$1905+0400$  &	     \phn \phn  $-3.80(18)  \phn              $ &      \phn $-7.3(4) \phn\phn\phn    $ &      \phn $-8.23	$ &      \phn \phn $-0.02 $ &  $6$\\   
J$1949+3106$  &	     \phn \phn  $-2.94(6)   \phn\phn          $ &      \phn $-5.17(8) \phn\phn       $ &      \phn $-5.95	$ &      \phn \phn $-0.14 $ &  $14$\\    
B$1957+20  $  &      \phn       $ -16.0(5)  \phn\phn\phn      $ &           $-25.8(6) \phn\phn\phn   $ &           $-30.36	$ &      \phn \phn $-0.03 $ &  $15$\\      
J$2019+2425$  &	     \phn \phn  $-9.41(12)  \phn              $ &           $-20.60(15) \phn         $ &           $-22.32	$ &      \phn \phn $-3.86 $ &  $16$\\     
J$2033+1734$  &	     \phn \phn  $-5.94(17)  \phn              $ &           $-11.0(3) \phn\phn\phn   $ &           $-12.42	$ &      \phn \phn $-1.43 $ &  $17$\\     
J$2124-3358$  &	     \phn       $-14.15(8)  \phn\phn          $ &           $-49.9(25)  \phn\phn     $ &           $-51.04	$ & \phs \phn \phn $9.06  $ &  $4$\\      
J$2129-5721$  &	\phs \phn \phn  $9.35(1)    \phn\phn          $ &      \phn $-9.47(1) \phn\phn       $ &           $-11.99	$ &      \phn \phn $-5.77 $ &  $5$\\    
J$2322+2057$  &	     \phn       $-17\phd(2) \phn\phn\phn\phn  $ &           $-18\phd(3) \phn\phn\phn\phn$ &        $-22.91	$ &      \phn \phn $-9.40  $ &  $18$
\enddata
\tablecomments{(1) \cite{Kerr2012}; (2) \cite{Du2014}; (3) \cite{Deller2008}; (4) \cite{Fermi2013}; (5) \cite{Verbiest2009}; (6) \cite{Gonzalez2011}; (7) \cite{Lazaridis2009}; (8) \cite{Ng2014}; (9) \cite{Deller2012}; (10) \cite{Konacki2003}; (11) \cite{Freire2012}; (12) \cite{Barr2013}; (13) \cite{Hou2014}; (14) \cite{Deneva2012}; (15) \cite{Arzoumanian1994}; (16) \cite{Nice2001}; (17) \cite{Splaver2004}; (18) \cite{Nice1995}}
\end{deluxetable*}

\clearpage

\makeatletter{}\begin{deluxetable}{lcc}
\centering
\tablewidth{0pt}
\tablecaption{Non-NANOGrav Parallax Measurements used in Velocity Analysis\label{tab:px_nonNANOGrav}}
\tablehead{\colhead{PSR} & \colhead{Parallax (mas)} & \colhead{Reference}}
\startdata
J$0437-4715$ & $6.40(5)$         & $1$ \\
J$0636+5129$ & $4.9(6) $\phn     & $2$ \\
J$1012+5307$ & $1.2(3) $\phn     & $3$ \\
J$1017-7156$ & $4\phd(1)  $\phn\phn & $4$ \\
J$1023+0038$ & $0.73(2)$         & $5$ \\
B$1257+12$   & $1.3(4) $\phn     & $6$ \\
J$1738+0333$ & $0.68(5)$         & $7$ \\
J$2124-3358$ & $3.1(6) $\phn     & $8$ 
\enddata
\tablecomments{(1) \cite{Deller2008}; (2) \cite{Stovall2014}; (3) \cite{Lazaridis2009}; (4) \cite{Ng2014}; \cite{Deller2012}; (6) \cite{Wolszczan2000}; (7) \cite{Freire2012}; (8) \cite{Verbiest2009}}
\end{deluxetable}

\clearpage

\makeatletter{}\begin{deluxetable}{lcc}
\centering
\tablewidth{0pt}
\tablecaption{Line-of-sight Velocities Used in our Analysis\label{tab:los_velocity}}
\tablehead{\colhead{PSR} & \colhead{LOS Velocity (km s$^{-1}$)} & \colhead{Reference}}
\startdata
J$1012+5307$ & \phs$44.0 \pm 8.0$ & 1\\
J$1903+0327$ & \phs$42.1 \pm 2.5$ & 2\\
B$1957+20  $ & \phs$85.0 \pm 5.0$ & 3
\enddata
\tablecomments{(1) \cite{Callanan1998}; (2) \cite{Khargharia2012}; (3) \cite{vanKerkwijk2011}}
\end{deluxetable}

\clearpage

\makeatletter{}\begin{deluxetable}{lccc}
\centering
\tablewidth{0pt}
\tabletypesize{\small}
\tablecaption{Galactic Components of Pulsar Velocities
\label{tab:velocity}}
\tablehead{
\colhead{PSR} & \colhead{Radial velocity} & \colhead{Azimuthal velocity} & \colhead{z velocity}
\\
              & \colhead{(km s$^{-1}$)  } & \colhead{(km s$^{-1}$)} & \colhead{(km s$^{-1}$)}
}
\startdata
\multicolumn{4}{c}{NANOGrav Millisecond Pulsars}\\[2pt]
\tableline
\rule{0pt}{10pt}J$0023+0923$ &          \phn $ -24.51  $  & \nodata              &                \nodata     \\
J$0613-0200$ &               \nodata      & \nodata              &               $  -13.54 $  \\
J$0645+5158$ &               \nodata      & \phs\phn    $37.93$  &                \nodata     \\
J$1024-0719$ &      \phn\phn $  -7.02  $  & \nodata              &                \nodata     \\
J$1600-3053$ &               \nodata      & \nodata              &          \phn $  -63.06 $  \\
J$1614-2230$ &               \nodata      & \phn       $-46.86$  &                \nodata     \\
J$1643-1224$ &               \nodata      & \phn       $-64.96$  &                \nodata     \\
J$1744-1134$ &               \nodata      & \phs\phn\phn $6.35$  &          \phn $  -34.56 $  \\
J$1853+1303$ &               \nodata      & \nodata              &  \phs\phn\phn $    7.89 $  \\
B$1855+09 $  &               \nodata      & \nodata              &  \phs\phn\phn $    6.47 $  \\
J$1903+0327$ &         \phn  $-30.61$     & \phn       $-37.44$  &   \phn        $  -11.44 $  \\
J$1909-3744$ &               \nodata      &           $-164.88$  &      \phn\phn $   -4.69 $  \\
J$1918-0642$ &               \nodata      & \nodata              &  \phs    \phn $   23.21 $  \\
J$1923+2515$ &               \nodata      & \nodata              &      \phn\phn $   -0.62 $  \\
B$1937+21 $  &      \phn\phn $  89.16  $  & \nodata              &  \phs\phn\phn $    2.86 $  \\
J$1944+0907$ &               \nodata      & \nodata              &               $ -197.68 $  \\
J$2043+1711$ &          \phn $ -43.41  $  & \nodata              &      \phn\phn $   -2.47 $  \\
J$2317+1439$ &  \phs\phn $      16.01  $  & \nodata              &                \nodata     \\
\cutinhead{Other measurements used in this work} 
J$0101-6422$ &      \phn\phn $  -9.21  $  & \nodata         &                 \nodata    \\
J$0218+4232$ &               \nodata      & \nodata         &       \phn\phn $   -4.19 $ \\
J$0437-4715$ &  \phs    \phn $  23.32  $  & \nodata         &                 \nodata    \\
J$0610-2100$ &               \nodata      & \nodata              &          \phs $  229.00 $  \\
J$0711-6830$ &  \phs    \phn $  31.77  $  & \nodata         &                 \nodata    \\
J$0751+1807$ &               \nodata      & \phs    $28.74$ &                 \nodata    \\
J$1012+5307$ &  \phs\phn\phn $3.15     $  & \phs    $72.16$ &   \phs    \phn $  74.74  $ \\
J$1017-7156$ &      \phn\phn $  -9.35  $  &         \nodata &   \phs\phn\phn $    7.69 $ \\
J$1045-4509$ &  \phs\phn\phn $   2.10  $  & \nodata         &   \phs    \phn $   25.41 $ \\
J$1125-5825$ &  \phs    \phn $  38.11  $  & \nodata         &       \phn\phn $   -4.14 $ \\
B$1257+12  $ &  \phs\phn\phn $   7.21  $  & \phn    $-4.92$ &                 \nodata    \\ 
J$1446-4701$ &               \nodata      & \nodata         &   \phs\phn\phn $    2.35 $ \\ 
J$1603-7202$ &               \nodata      & \nodata         &       \phn\phn $  -8.65  $ \\ 
J$1738+0333$ &               \nodata      & \nodata         &           \phn $  -18.26 $ \\
J$1745+1017$ &               \nodata      & \nodata         &           \phn $  -35.97 $ \\
J$1843-1113$ &               \nodata      & \nodata         &   \phs    \phn $   12.36 $ \\
J$1905+0400$ &               \nodata      & \nodata         &   \phs\phn\phn $    6.87 $ \\
J$1949+3106$ &               \nodata      & \nodata         &   \phs\phn\phn $    2.35 $ \\
B$1957+20  $ &               $-305.27  $  & \phs \phn $0.27$&   \phn\phn     $  -0.07  $ \\
J$2019+2425$ &               $-118.90  $  & \nodata         &           \phn $  -19.82 $ \\
J$2033+1734$ &          \phn $ -66.22  $  & \nodata         &       \phn\phn $   -7.39 $ \\
J$2124-3358$ &               \nodata      &        $-41.93$ &                 \nodata    \\
J$2129-5721$ &               \nodata      &        $-37.50$ &                 \nodata    \\
J$2322+2057$ &          \phn $ -60.98  $  & \nodata         &                 \nodata         
\enddata
\tablecomments{Velocities listed here include only those components whose vector is within $20^{\circ}$ of perpendicular to the line-of-sight vector.} 
\end{deluxetable}

\clearpage

\makeatletter{}\begin{deluxetable}{lrrrrc}
\centering
\tablewidth{0pt}
\tabletypesize{\small}
\tablecaption{PSR~J1024$-$0719 Comparison of Spin-down and Proper Motion Measurements\tablenotemark{a,b}
\label{tab:1024_previous}
}
\tablehead{
                    & \colhead{$\pdot$}      &   \colhead{$\mu_{\alpha}$}     &   \colhead{$\mu_{\delta}$}     & Data Span \\
                    & \colhead{$(10^{-20}$)} &   \colhead{(mas yr$^{-1}$)}    &   \colhead{(mas yr$^{-1})$}    & (Years) 
}
\startdata
\cite{Hotan2006}    &   1.853(6)\phn        &          -34.9(4)               &  -47(1) \phn           & 2003.0-2005.3 \\
\cite{Verbiest2009} &   1.852(8)\phn        &          -35.2(2)               &  -48.2(3)              & 1996.1-2008.2 \\
NANOGrav Nine-year Release (this work)  &   1.8551(1)  &  -34.2(1)            &  -48.0(3)              & 2009.8-2013.8 
\enddata
\tablenotetext{a}{Numbers in parentheses are 1$\sigma$ uncertainties in the last digit quoted.}
\tablenotetext{b}{Parameters for this pulsar were also reported by \cite{Toscano1999}, but they 
disagree with the values reported here, despite an overlap in data span between that work
and \cite{Verbiest2009}.  Based on the consistency of other measurements, we have excluded
the \cite{Toscano1999} parameters from our analysis.}
\end{deluxetable}

\clearpage

\makeatletter{}\begin{deluxetable}{lrrrr}
\centering
\tablewidth{0pt}
\tabletypesize{\small}
\tablecaption{PSR~J1024$-$0719 Parallax Measurement Trials
\label{tab:1024_px}
}
\tablehead{
\colhead{Trial} &    \colhead{$\varpi$ (mas)}   &   \colhead{$\chi^2$}    &   \colhead{$n_{\rm dof}$}    &   \colhead{$\chi^2/n_{\rm dof}$}
\\
}
\startdata
\multicolumn{5}{c}{Standard solution, full data set}\\[2pt]
\tableline
\rule{0pt}{10pt}Standard solution    &   $0.63   \pm 0.29$    &   4762.18     &     4766           &  0.9992
\\
\cutinhead{Data subsets}
\rule{0pt}{10pt}GUPPI data only      &   $0.62   \pm 0.29$    &   4686.30     &     4690           &  0.9992
\\
2009.8-2011.8 data only &   $0.71   \pm 0.64$    &   1734.74     &     1697           &  1.0222
\\
2011.8-2013.8 data only &   $0.81   \pm 0.38$    &   2957.44     &     3005           &  0.9842
\\
\cutinhead{Extra timing noise terms}
\rule{0pt}{10pt}     Fit $f_2$       &   $0.72   \pm 0.29$    &   4759.53     &     4765           &  0.9988
\\
     Fit $f_2$, $f_3$&   $0.75   \pm 0.29$    &   4758.32     &     4764           &  0.9988
\\
     Fit $f_2,\dots,f_6$ &   $0.77   \pm 0.30$    &   4756.11     &     4761           &  0.9990
\\
\cutinhead{Modified dispersion measure models}
\rule{0pt}{10pt}Linear DM; $n_{0,solar}=0~{\rm cm}^3$   &   $0.74   \pm 0.12$    &   4921.10     &     4817           &  1.0216
\\
Linear DM; $n_{0,solar}=10~{\rm cm}^3$  &   $1.11   \pm 0.12$    &   4924.40     &     4817           &  1.0222

\enddata
\end{deluxetable}

\clearpage

\makeatletter{}

\clearpage

\makeatletter{}

\clearpage

\makeatletter{}

\clearpage

\makeatletter{}

\clearpage

\makeatletter{}

\clearpage

\makeatletter{}

\clearpage

\makeatletter{}

\end{document}